\documentclass{aa}
\usepackage{txfonts}

\usepackage{amsmath}
\usepackage{graphicx}
\usepackage{multirow}
\usepackage{array}
\usepackage{natbib}
\bibpunct{(}{)}{;}{a}{}{,}

\newcommand{\beq}{\begin{equation}}
\newcommand{\bea}{\begin{eqnarray}}
\newcommand{\eeq}{\end{equation}}
\newcommand{\eea}{\end{eqnarray}}
\newcommand{\p}{\partial}
\newcommand{\BB}{{\bf B}}
\newcommand{\UU}{{\bf V}}

\begin{document}
   \title{Postshock turbulence and diffusive shock acceleration\\ in young supernova remnants}
   
  \author{A. Marcowith\inst{1} \and F. Casse\inst{2}}

\titlerunning{Postshock turbulence in young supernova remnants}
\authorrunning{Marcowith \& Casse} 

\institute{
  Laboratoire de Physique Th\'eorique et d'Astroparticules (LPTA)\\ UMR
  5207 CNRS - Universit\'e Montpellier II\\ 13 place E. Bataillon F-34095
  Montpellier Cedex 5, France\\ \email{Alexandre.Marcowith@LPTA.in2p3.fr}\\
\and
Laboratoire AstroParticule \& Cosmologie (APC)\\ UMR 7164 CNRS -
  Universit\'e Paris Diderot\\ 10, rue Alice Domon et L\'eonie Duquet
  F-75205 Paris Cedex 13\\  \email{fcasse@apc.univ-paris7.fr}}

   \date{Received ...; accepted ...}

   \abstract{Thin X-ray filaments are observed in the vicinity of young supernova remnants (SNR) blast waves.  Identifying the process that creates these filaments would provide direct insight into the particle acceleration occurring within SNR and in particular the cosmic ray yield.}  
 {We investigate magnetic amplification in the upstream medium of a SNR blast wave through both resonant and non-resonant regimes of the streaming instability. We attempt to understand more clearly of the diffusive shock acceleration (DSA) efficiency by considering various relaxation processes of the magnetic fluctuations in the downstream medium. Multiwavelength radiative signatures originating in the SNR shock wave are used to test various downstream turbulence relaxation models.}
{Analytical and numerical calculations that couple stochastic differential equation schemes with 1D spherical magnetohydrodynamics simulations are used to investigate, in the context of test particles, turbulence evolution in both the forshock and post-shock regions. Stochastic second-order Fermi acceleration induced by resonant modes, magnetic field relaxation and amplification, and turbulence compression at the shock front are considered to model the multiwavelength filaments produced in SNRs. The $\gamma$-ray emission is also considered in terms of inverse Compton mechanism.}
{We confirm the result of Parizot and collaborators that the maximum CR energies should not go well beyond PeV energies in young SNRs where X-ray filaments are observed. To reproduce observational data, we derive an upper limit to the magnetic field amplitude and so ensure that stochastic particle reacceleration remains inefficient. Considering various magnetic relaxation processes, we then infer two necessary conditions to achieve efficient acceleration and X-ray filaments in SNRs: (1) the turbulence must fulfil the inequality $2-\beta-\delta_{\rm d} \ge 0$; where $\beta$ is the turbulence spectral index and $\delta_d$ is the relaxation length energy power-law index; (2) the typical relaxation length must be of the order the X-ray rim size. We find that Alv\'enic/fast magnetosonic mode damping fulfils all conditions; while non-linear Kolmogorov damping does not. By confronting previous relaxation processes with observational data, we deducte that among our SNR sample, data for the older ones (SN1006 and G347.3-0.5) does not comply with all conditions, which means that their X-ray filaments are probably controlled by radiative losses. The younger SNRs, Cassiopeia A, Tycho, and Kepler pass all tests and we infer that the downstream magnetic field amplitude is in the range of 200-300 $\mu$ Gauss.} 
{}
\maketitle             
%
\section{Introduction} 
\label{S:Intro}
Chandra high-angular resolution X-ray observations of young supernova remnants (SNR) such as Cassiopeia A, Kepler or Tycho, have detected
very thin X-ray filaments, which are probably associated with the supernova (SN) forward shock expanding into the interstellar medium (ISM) \citep{Gotthelfetal01,Hwangetal02,Rhoetal02,Uchiyamaetal03,Cassamchenaietal04,Bambaetal05a,Bambaetal05b,Cassamchenaietal07}. The physical properties
of these filaments were reviewed by \cite{Vink03}, \cite{Vink04}, \cite{Weisskopf05}, \cite{Ballet06}, \cite{Parizotetal06}, \cite{Bambaetal06}, and \cite{Berezhko08}. These  filaments are believed to be produced by synchrotron radiation emitted by TeV electrons. Rim-like filaments are usually of a few arcsecond in angular size as reported in \cite{Parizotetal06}. Their true width, however has to be inferred from  deprojection calculations by taking into account the curvature of the remnant \citep{Berezhkoetal03a, Ballet06}.  This size should depend on the magnetic field strength, local fluid properties (the shock velocity and compression ratio), and  the relativistic electron diffusion regime.

Measurements of the X-ray rim size inferred a lower limit to the magnetic field located downstream from the shock front. Typical field strengths of two orders of magnitude above the standard ISM values $B_{\infty}$  were reported by  e.g., \cite{Berezhkoetal03a}, \cite{Vink04}, \cite{Voelketal05}, \cite{Parizotetal06}, and \cite{Berezhko08}. \cite{Vink04} showed that  advective and diffusive transport also contributes to the filament extension at high energy close to the electron cut-off. The aforementioned constraints infer a value of the electron spatial diffusion coefficient that is a few times higher than the Bohm limit in the downstream region from the shock \footnote{The Bohm diffusion coefficient is obtained when the particle mean free path is equal to its Larmor radius $r_{\rm L} = E/ZeB$, i.e., $D_{\rm Bohm} = r_{\rm L} c/3$.}. These results support the standard scenario of diffusive shock acceleration (DSA) in SNRs and imply that strong magnetic field amplification occurs at the shock precursor. However,  Chandra observations have been obtained for a limited frequency range. Thus, diffusion regimes differing from that of the Bohm diffusion cannot be ruled out by these sole observations \citep{Marcowithetal06}. For instance, alternative diffusion regimes may affect high energy particle transport and modify the way in which the synchrotron spectrum cut-off is reconstructed from the extrapolation of the radio spectrum \citep{Zira07}.  However, the hard X-ray detection of SNR RXJ1713-3946.5 by Suzaku \citep{Takahashietal08} supports a cut-off spectrum in agreement with a Bohm-like diffusion regime.

The origin of the magnetic fluctuations sustaining the diffusive behavior of non-thermal particles remains widely debated. One possibility is that the turbulent magnetic field is generated by the relativistic particles themselves by means of their streaming motion ahead of the shock front \citep{Bell01}. The field amplification has strong implications for the physics of cosmic-ray (CR) acceleration at SNR shocks. For instance, a calculation including the effect of non-linear turbulence transfer concluded that proton acceleration is possible up to the CR spectrum knee at $\sim 3 \times 10^{15} eV$.  This calculation was performed in the most extreme shock velocity regimes, particularly for SNRs propagating in a hot interstellar medium free of ion-neutral wave damping \citep{Ptuskin03}. \cite{Bell04} discussed a non-resonant regime of streaming instability that can generate a very strong turbulent magnetic field (and boost the CR maximum energy) readily at the very early stage of the SNR free expansion phase. \cite{Diamond07} and \cite{Pelletieretal06} also highlighted the importance of  determining the saturation level of the magnetic fluctuations, which was partially discarded in the previous work. \cite{Pelletieretal06} demonstrated that both resonant and non-resonant regimes of the streaming instability have to be considered simultaneously to fix the magnetic field spectrum and strength at the shock front. In fast shocks, the non-resonant instability dominates the magnetic field generation, the level of fluctuation at the shock being found to be similar to the value derived by \cite{Bell04}. The resonant instability dominates in slower shock regimes.  The turbulence generated upstream may then relax downstream from the shock front, limitating of the spatial extent of the non-thermal particle journey \citep{Pohletal05}. This possibility has not yet been completely taken into account in the DSA process and the corresponding maximum energy reachable by relativistic particles. This issue is investigated in a dedicated section of the present article.
We note that the problem of the maximum CR energy was addressed by \cite{Zira08} using a semi-analytical approach to the non-resonant streaming mode generation. The authors identified the maximum CR  energy, between the two confinement limits, expected for a standard ISM medium or a completely amplified magnetic field. One should keep in mind that several effects can alterate these conclusions such as the propagation into a partially ionised medium \citep{Bykov05, Reville07}, thermal effects in the dispersion relation of the non-resonant instability \citep{Reville08}, or a back reaction on the CR current \citep{Riquelme09}.

Although disputed (see discussions in \cite{Katz08} and \cite{Morlinoetal08}), the production of relativistic hadrons in SNRs is consistent with detection of a few TeV $\gamma$-ray emitting SNRs in the Galactic plane by the HESS telescope. This $\gamma$-ray emission may favor the interaction of relativistic hadrons with a dense molecular cloud leading to the Compton upscattering of low energy photons \citep{Aharonianetal04, Aharonianetal06, Albertetal07}. Nevertheless, more observations are mandatory before drawing any firm conclusion about this important issue.

The present article investigates DSA processes involving magnetic field amplification and relaxation. The paper considers the effect of shock acceleration, spatial variation in the magnetic field (and the corresponding diffusion coefficient), the possibility of finite diffusive extension zones, and the effect of stochastic Fermi acceleration by the electromagnetic fluctuations generated in the shock precursor. This modelling is performed by means of numerical calculations. The numerical scheme is based upon the stochastic differential equations (SDE) and is described in Appendix \ref{S:Appc}.  Sect. \ref{S:Upa} presents the general framework adopted in this article. In particular, it investigates the conditions required to develop turbulence upstream from the shock, as expected from the non-linear evolution of the various regimes of the streaming instability. Sect. \ref{S:Cmfp} and \ref{S:DoMF} investigate the impact of post-shock turbulence upon particle acceleration. Sect. \ref{S:Cmfp} dealing with advected downstream turbulence and sect. \ref{S:DoMF} refering to a downstream relaxing turbulence. All calculations are then compared with those for a sample of young SNRs presented in \cite{Parizotetal06} already.\\
Table (\ref{T:tab0}) summarises the notations used in this article (the section where the parameter is reported at first is also indicated).

\section{Upstream turbulence generation and accelerated particle diffusion}
\label{S:Upa}

Highly turbulent supernova shocks involve several complex processes that modify the standard DSA model at some stage of the SNR evolution. 
In the upstream region, the properties of the turbulence are controlled by the fastest growing instability and its saturation mode
\citep{Pelletieretal06}. The diffusion regime strongly depends on the competition between the wave growth and the energy transfer
to other scales provoked by non-linear cascades \citep{Marcowithetal06}. The turbulence is then compressed at the shock-front, i.e., parallel modes (parallel to the shock normal) have  wavelengths that are shorter by a factor equal to the (sub)shock compression ratio. In the downstream region, the 
turbulence can either be relaxed \citep{Pohletal05} or amplified \citep{Pelletieretal06, Zira08}. The turbulent magnetic field coherence length may also vary with the distance to the shock, which can be modelled using self-similar solutions \citep{Katzetal07}.

Section \ref{S:UpMF} summarises the properties of the two regimes (both resonant and non-resonant) of the streaming instability as well as the magnetic field profiles inserted into the coupled SDE-magnetohydrodynamics (MHD) numerical calculations. In sect. \ref{S:Dup}, we derive the general form of the diffusion coefficient. Finally, sect. \ref{S:Pdif} displays the general expression of the particle distribution function, at the shock front, expected in the case of spatially varying diffusive zones. The various expressions derived in this section will be used in sect. \ref{S:Cmfp} and \ref{S:DoMF}.
\begin{table*}
\begin{tabular}[t]{|cl|}
\hline
\multirow{1}{40mm}{\bf Turbulence parameters} &
\begin{tabular}[t]{|cl|} 
\firsthline
\multirow{1}{3mm}{$\beta$} & One D power-law spectral index of the  turbulence spectrum [Eq.(\ref{Eq:nus})] \\  \hline
\multirow{1}{3mm}{$\eta_{\rm T}$} & Level of magnetic fluctuations  with respect to the mean ISM magnetic field  [Eq.(\ref{Eq:nus})]  \\ \hline
\multirow{1}{3mm}{$\phi$} &  Logarithm of the ratio of the maximum momentum to the injection momentum [Eq.(\ref{Eq:BNRB0})] \\ \hline
\multirow{1}{3mm}{$\lambda_{\rm max}$} & Longest wavelength of the magnetic turbulence spectrum (Sect. \ref{S:Dup}) \\ \hline
\multirow{1}{3mm}{$\ell_{\rm coh}$} & Coherence length of the magnetic fluctuations (Sect. \ref{S:Dup}) \\ \hline
\multirow{1}{3mm}{$\sigma$} & Normalisation factor entering  the turbulent spectrum (Sect. \ref{S:Dup}) \\ \hline
\multirow{2}{3mm}{$\delta_{\rm u/d}$} & Power-law energy dependance index of the relaxation lengths either up- \\ & or downstream (Sect. \ref{S:DoMF} and [Eq.(\ref{Eq:Elld}]) \\ \hline
\multirow{1}{3mm}{$H$} & Ratio of the upstream to the downstream diffusion coefficient at the shock front [Eq.(\ref{Eq:D})]\\ \hline
\multirow{1}{3mm}{$\delta_{\rm B}$} & Ratio of the resonant to the non-resonant magnetic field strength at the shock front [Eq.(\ref{Eq:BRBNR})]    \\\lasthline
\end{tabular} \\ \hline
\multirow{1}{40mm}{\bf Relativistic particle parameters} &
\begin{tabular}[t]{|cl|}
\firsthline 
\multirow{1}{8mm}{$\xi_{\rm CR}$} & Ratio of the CR pressure to the shock dynamical pressure [Eq.(\ref{Eq:BRBNR})] \\ \hline
\multirow{1}{8mm}{$r_L$} &  Larmor radius of a particle  (defined using resonant magnetic field) \\\hline
\multirow{1}{8mm}{$\rho$} & Ratio of the particle Larmor radius to $\lambda_{\rm max}/2\pi$  (also called reduced rigidity, see sect. \ref{S:Dup})\\ \hline
\multirow{1}{8mm}{$E_{\rm CR-max}$} & Maximal cosmic ray energy  (Sect. \ref{S:Dup}) \\ \hline
\multirow{1}{8mm}{$E_{e-max}$} & Maximal electron energy  (Sect. \ref{S:Adva})\\ \hline
\multirow{1}{8mm}{$E_{\rm \gamma-cut}$} & Cut-off synchrotron photon energy  emitted by electrons at $E_{\rm e-max}$ (Sect. \ref{S:Adva}) \\ \hline
\multirow{1}{8mm}{$E_{\rm CR-min}$} & Injection energy of the cosmic rays   (Sect. \ref{S:TDD}) \\ \hline
\multirow{1}{8mm}{$E_{e-obs}$}& Energy of the electrons producing the observed X-ray filaments (Sect. \ref{S:Mfrel}) \\\lasthline
\end{tabular} \\ \hline
 \multirow{1}{40mm}{\bf SNRs parameters} &
 \begin{tabular}[t]{|cl|}
\firsthline 
 \multirow{1}{12mm}{$V_{\rm sh,4}$} & Velocity of the SNR shock wave (in $10^4$ km/s unit)\\ \hline
 \multirow{2}{12mm}{$B_{\rm d/u,-4}$} & Magnetic field amplitude at the shock front respectively in the \\ & down- and upstream medium (in $10^{-4}$ Gauss unit)\\ \hline
 \multirow{1}{12mm}{$r_{\rm B}, r_{\rm sub}, r_{\rm tot}$} & Magnetic, sub-shock, and total shock compression ratios (Sect. \ref{S:Adva}) \\ \hline
 \multirow{1}{12mm}{$\Delta R_{\rm X,-2}$} & X-ray filament deprojected width (in $10^{-2}$ parsec unit, sect. \ref{S:Mfrel}) \\\lasthline
\end{tabular} \\ \hline
\multirow{1}{40mm}{\bf Equation parameters} &
 \begin{tabular}[t]{|cl|}
\firsthline 
\multirow{1}{12mm}{${\rm y}(r)$} & $3 r^2/(r-1)$ [Eq.(19)] \\ \hline  
\multirow{1}{12mm}{${\rm K}(r,\beta)$} & $q(\beta) \times (H(r,\beta)/r+1)$ [Eq.(36)] \\ \hline 
\multirow{1}{12mm}{${\rm f}_{\rm sync}$} & $H(r,\beta)+r / H(r,\beta)/r_{\rm B}^2 +r$ [Eq.(39)] \\ \hline
\multirow{1}{12mm}{$g(r)$}  & $3/(r-1) \times (H(r,\beta)/r + 1)$  [Eq.(40)] \\ \hline 
\multirow{1}{12mm}{$C(\delta_d)$} &  $(E_{\rm e-max}/E_{e-obs})^{\delta_d}$  [Eq.(41)] \\ \lasthline 
\end{tabular} \\\hline
\end{tabular}
\caption{Summary of the notations used in this article to denote the various physical quantities and parameters involved in our description of high energy particle yield in supernova remnants (SNR). }
\label{T:tab0}  
\end{table*}
\subsection{Cosmic-ray streaming instabilities}
\label{S:UpMF}
The streaming instability which is provoked by the superalfvenic motion of accelerated energetic particles, generates magnetic fluctuations over a large interval of wave numbers. The resonant instability involves wave-particle interaction on wave scales of the order of the particle gyro-radius $r_{\rm L}$ \citep{Skilling75, Bell01}. The non-resonant regime was adapted to SNR shock waves by \cite{Bell04} (see also \cite{Pelletieretal06}, \cite{Zira08}, and \cite{Amato09} for further details). 
The non-resonant waves are produced, at least in the linear growth phase of the instability, on  scales much smaller than $r_{\rm L}$. However, the ability of the instability to both enter deeply into the non-linear regime and saturate at a magnetic field level $\delta B \gg B_{\rm \infty}$ remains debated \citep{Reville08, Niemiecetal08, Riquelme09}. In the next paragraph, we summarise the main properties of the wave modes generated by the non-resonant streaming instability (sect. (\ref{S:MFnr})). We then present the characteristics  of the resonant regime in sect. (\ref{S:MFr}).
\subsubsection{The non-resonant regime}
\label{S:MFnr}
In the linear phase, the most rapidly growing waves have large wave numbers \citep{Bell04} given by
\beq
k \le k_{\rm c} = {j_{\rm cr} B_{\infty} \over \rho_{\infty} V_{\rm a\infty}^2 c} \ ,
\eeq
where $j_{\rm cr} = n_{\rm cr}e V_{\rm sh}$ is the current produced by the cosmic rays ahead of the shock wave, $n_{\rm CR}$ 
is the CR density, and $V_{\rm sh}$ is the shock velocity measured in the upstream restframe. 

The wave number corresponding to the maximal growth rate $\gamma_{\rm max} = k_{\rm up} V_{\rm a\infty}$ is
\beq
k_{\rm up} = {k_{\rm c} \over 2} = {n_{\rm cr} \over n_{\rm \infty}} \times \Omega_{\rm cp} \times {V_{\rm sh} \over 
2 V_{\rm a\infty}^2} \ ,
\eeq
where $n_{\infty}$, $\Omega_{\rm cp} = eB_{\rm \infty}/(m_{\rm p}c)$, and $V_{\rm a\infty}=B_{\rm \infty}/\sqrt{4\pi n_{\rm \infty}}$ 
are, respectively, the density, the cyclotron frequency, and the Alfv\'en velocity in the ISM \footnote{The density $n_{\infty}$ is usually the 
ion density, but when the coupling between ion and neutrals is effective it must also involve the density of neutrals}.

MHD calculations \citep{Bell04, Zira08} have shown that beyond an exponential growth  phase located on
typical scale of
\[
x_{\rm g} = V_{\rm sh}/\gamma_{\rm CR-max} 
\]
from the shock, the instability enters the non-linear regime. The magnetic fluctuations are redistributed on larger scales, while the turbulence level evolves in a linear way. \cite{Bell04} and \cite{Pelletieretal06} discussed several saturation processes that all lead to an energy transfer from the dominant wavelength towards long wavelengths (see discussion in \cite{Riquelme09}).  One may then expect the coherence length of the 
turbulence to be transferred from a scale $\ell_{\rm coh-L}$, where $k_{\rm max}^{-1} \le \ell_{\rm coh-L} \ll \bar{r}_{\rm L-CR-max}$, 
to a scale $\ell_{\rm coh-NL} < \bar{r}_{\rm L-CR-max}$ where $\bar{r}_{\rm L-CR-max} = r_{\rm L-CR-max} \times 
B_{\rm \infty}/\bar{B}$ is the renormalised maximum energy CR gyro-radius in the amplified magnetic field $\bar{B}$. Resonant modes have a harder spectrum \citep{Pelletieretal06} hence contribute to an increase in the coherence length of the turbulence (see sect. \ref{S:MFr}). So, hereafter, we consider both regimes producing a turbulence with a coherence scale close to $\bar{r}_{\rm L-CR-max}$,  i.e., we neglect the extension of the upstream region where the non-resonant instability is in the linear regime (see sect. \ref{S:Dup}).

Another important property of non-resonant modes is that they have non-vanishing helicity \citep{Pelletieretal06}. These modes are mostly proton-induced and have a right-handed polarisation with respect to the mean magnetic field far upstream. This non-zero helicity may be the origin of
additional amplification in the downstream medium, where the total magnetic field can eventually reach values close to the equipartition with the kinetic energy of the thermal gas. 
\subsubsection{The resonant regime}
\label{S:MFr}
The resonant regime develops simultaneously with the non-resonant regime \citep{Pelletieretal06} and {\it cannot be discarded}. The total 
amplification factor of the magnetic field $A_{\rm tot}^2 = B_{\rm tot}^2/B_{\rm \infty}^2$ at a distance $x$ from the shock front is a combinaition of both  non-resonant and resonant contributions, namely $A^2_{\rm tot}(x)=A^2_{\rm NR}(x)+A^2_{\rm R}(x)$. The exact spatial 
dependence of $A_{\rm R}(x)$ is derived in Appendix \ref{S:Appa} for completeness. It is found that a good approximation is $A_{\rm R} \propto A_{\rm NR}^{1/2}$. \\
To quantify the previous assertion, we parametrise the contribution of each instability regime. \cite{Pelletieretal06} argued that the 
shock velocity is the main controlling factor of each contribution. This dependence can be inferred from Eq.(\ref{Eq:Arf}). By comparing the respective saturation values of each regime, one finds that
\beq
\label{Eq:BRBNR}
\frac{B_{\rm R}(x=0)}{B_{\rm NR}(x=0)}= \delta_{\rm B} = \left(\frac{\xi_{\rm CR} c}{V_{\rm sh}}\right)^{1/4} \ ,
\eeq
while
\beq
\label{Eq:BNRB0}
\frac{B_{\rm NR}(x=0)}{B_{\rm \infty}} = \delta_{\rm B}^{-6} \times \left(\frac{3 c^2 \xi_{\rm CR}^4}{\phi  V_{\rm A\infty}^2}\right)^{1/2} \ .
\eeq
The level of magnetic fluctuations at the shock front given by Eq. (\ref{Eq:BRBNR}) and (\ref{Eq:BNRB0}) is controlled by both $\delta_{\rm B}$
and the fraction $ \xi_{\rm CR}$ of the SNR dynamical pressure transferred into the CR. The parameter $\phi = \log(p_{\rm max}/p_{\rm inj})$ is
the logarithm of the ratio of the maximum to the injection momentum and is approximately between  $15$ and $16$.

As a fiducial example, we assume that $\xi_{\rm CR}=0.2$, $B_{\rm \infty} = 4 \ \rm{\mu Gauss}$, and that the ion density as$n_{\rm i} = 0.7 \ \rm{cm^{-3}}$. We then identify three distinct domains:
\begin{enumerate}
\item $\delta_{\rm B} \ge 3$ (corresponding to $V_{\rm sh} \le \rm{c/400}$) in which the magnetic field amplification provided by the streaming instability is modest for slow shock velocities. 
\item $1$ $< \delta_{\rm B} < 3$ (corresponding to $\rm{c/400} < V_{\rm sh} < \rm{c/10}$): for which we get the ordering $B_{\rm R} \ge B_{\rm NR} > B_{\rm \infty}$ and, that the ratio 
 $B_{\rm R} / B_{\rm NR}$ does not exceed a factor $2$.
\item  $\delta_{\rm B} \le 1$ (corresponding to $V_{\rm sh} \ge \rm{c/10}$): for which the magnetic ordering becomes $B_{\rm NR} \ge B_{\rm R} > B_{\rm \infty}$. In that case, an upper limit velocity stands close to $c$. Beyond that limit, the amplification by the non-resonant instability is maximal. An accurate analysis is then necessary to compare the saturation
of the instability induced by both advection and non-linear effects \citep{Pelletieretal09}.
\end{enumerate}
\noindent Electrons and protons (or ions) moving in the forward or backward direction can resonate with either forward or backward modes. Efficient mode redistribution is expected to produce waves in 
both directions in the shock precursor (see the Appendix of \cite{Pelletieretal06} for further details).  We note that the interaction between resonant Alfv\`en waves and the shock produce  magnetic helicity that is different from either $+1$ or $-1$ and ensures that second order Fermi acceleration by the resonant modes is unavoidable in the downstream region \citep{Campeanu92, Vainio99}. This effect is discussed in sect. \ref{S:FII}.

\subsection{A note on the evolution of non-resonant modes}
\label{S:NRmod}
Non-resonant modes are purely growing modes of null frequency, at least in the linear phase. They do not correspond to any normal mode of the plasma as 
in the case of the resonant regime. Consequently they are expected to be rapidly damped once the source term is quenched, i.e., at the shock front. The damping length
should be of the order of a few plasma skin depths. However, these modes also have a non-vanishing helicity \citep{Pelletieretal06, Zira08} (as we see in sect. \ref{S:TDD} ).
So a fraction of the turbulent spectrum can grow further downstream by means of dynamo action. At this point, the downstream evolution of the non-resonant spectrum is
unclear. In some conditions the combinaition of magnetic field compression and non-resonant mode damping at the shock front leads to a downstream magnetic
field that is weaker than the upstream field, especially in the very fast shock regime (regime 3. discussed in sect. \ref{S:MFr}). This is not the case for the SNR sample considered in this work as
the resonant modes tend to be dominant at the shock front. A complete investigation of this difficult issue would require a detailed investigation of the interstellar medium interaction with shocks to fix the ratio $B_{\rm R}/B_{\rm NR}$. For this reason, we assume hereafter that the downstream behaviour of the turbulence is dominated
by the resonant mode. However, even if  $B_{\rm R}/B_{\rm NR} > 1$ at the shock front, the fastest growing channel is the non-resonant one, which is important for the complete setting of the magnetic field turbulence in the upstream region. We acknowledge that this assumption weakens the analysis presented in the following sections and
consider this first work to be an attempt to isolate the main properties of the turbulence around a SNR shock.

\subsection{Upstream diffusion regimes}
\label{S:Dup}
As previously discussed, the most energetic CRs generate fluctuations at scales that are much smaller than $ r_{\rm L-CR-max}$. These particles experience small-scale turbulence exclusively in the unamplified magnetic field. Thus, the diffusion coefficient at maximum energy scales as $D(E_{\rm CR-max})=(r_{\rm L-CR-max}/\ell_{\rm coh})^2 \ell_{\rm coh}c$ (see below).
This allows us to compare $x_{\rm g}$ and $\ell_{\rm diff}(E_{\rm CR-max}) = D(E_{\rm CR-max})/V_{\rm sh}$, the diffusive length of the most energetic cosmic rays.  
One can then write \citep{Pelletieretal06}
\beq
x_{\rm g} = {2\phi \over 3 \xi_{\rm CR}} \times {P_{\rm B\infty} \over \rho_{\rm \infty} V_{\rm sh}^2} \times {V_{\rm sh} \over V_{\rm a\infty}} \times 
{\ell_{\rm coh} \over r_{\rm L-CR-max}} \times \ell_{\rm diff}(E_{\rm CR-max})  \ .
\eeq
We find that $x_{\rm g} \ll \ell_{\rm diff}(E_{\rm CR-max})$ in fast shocks ($V_{\rm sh} > 10^{-2} \rm{c}$) because $(P_{\rm B\infty}/\rho_{\rm \infty} V_{\rm sh}^2) \ V_{\rm sh}/V_{\rm a\infty} \sim V_{\rm a\infty}/V_{\rm sh}$. The following notations were used to derive the previous result:  the CR density is linked to the CR pressure by $n_{\rm CR}= 3 P_{\rm CR}/\phi p_* c$ and $p_*=p_{\rm CR-max}$ at a distance $x=\ell_{\rm diff}(E_{\rm CR-max})$ from the shock.  The parameter $\xi_{\rm CR}$ is probably between 0.1 and 0.3.

CRs and electrons of energy  lower than $E_{\rm CR-max}$, diffuse by means of a large-scale turbulence, their transport properties differing from those of most energetic CRs \citep{Zira08}. 
Whatever the turbulence level, the angular diffusion frequency (for a relativistic particle in an amplified field) can be estimated as (see \citet{Casseetal02}, their Eq.A22) :
\beq
\label{Eq:nus}
\nu_{\rm s} \simeq {\pi \over 3} \bar{r}_{\rm L}^{-2} \times (\beta-1) \times b \, c \, {\delta B^2 \over \bar{B}^2} \ ,
\eeq
where 
\beq
b= \ell_{\rm coh} \times \int_{k_{\rm min-NR} \ell_{\rm coh}}^{k_{\rm max-NR} \ell_{\rm coh}} d\ln(k) (k\ell_{\rm coh})^{-\beta}\label{Eq:func_b} \ .
\eeq
The turbulence spectrum is assumed to spread over the range $[k_{\rm max}^{-1}, k_{\rm min}^{-1}]$ with a 1D power-law spectral index $\beta$. 
If $\beta = 1$, the term $1/(\beta-1)$ has to be replaced by a factor $\sigma=\ln(k_{\rm max}/k_{\rm min})$. The corresponding spatial diffusion coefficient is by definition 
$D={c^2/3 \nu_{\rm s}}$. Its energy dependence is related to the development of the instability. In the linear phase 
(small scale turbulence), we recover the above expression for $\ell_{\rm diff}(E_{\rm CR-max})$. If $k_{\rm min-NR} \ell_{\rm coh}\simeq 1$, after introducing the level of turbulence $\eta_{\rm T}=\delta B^2/\bar{B}^2$, we find that
\beq
\label{Eq:D2}
D(E)={\beta \over \pi (\beta-1)} \times {\ell_{\rm coh} c \over \eta_{\rm T}} \times \left({\bar{r}_{\rm L} 
\over \ell_{\rm coh}}\right)^2 \ .
\eeq
In the non-linear phase (i.e., large-scale turbulence), we have $k_{\rm min-NR} \sim \bar{r}_{\rm L}$ and so 
\beq
\label{Eq:D2b}
D(E)={\beta \over \pi (\beta-1)} \times {\ell_{\rm coh} c \over \eta_{\rm T}} \times \left({\bar{r}_{\rm L} \over 
\ell_{\rm coh}}\right)^{2-\beta} \ .
\eeq
The results obtained by \cite{Casseetal02} can be recovered using $\ell_{\rm coh} = \rho_{\rm M} \lambda_{\rm max}/2\pi$ by
adopting a reduced rigidity of $\rho = 2\pi \bar{r}_{\rm L}(E)/\lambda_{\rm max}$. The length $\lambda_{\rm max} \simeq  20 \ell_{\rm coh}$ is the 
maximum scale of the turbulence and $\rho_{\rm M}$ is a number $\sim 0.2-0.3$. This latter number corresponds to the reduced rigidity at which the transition 
between the two diffusive regimes operates. For instance, assuming $\eta_{\rm T} \simeq 1$ and 
$\beta = 5/3$,  one finds that $D\simeq 2.2 D_{\rm Bohm}$ at $\bar{r}_{\rm L}=\ell_{\rm coh}$, which is consistent with the numerical solutions found by \cite{Casseetal02}. If $\beta = 1$, the energy independent ratio $D/D_{\rm Bohm} = 3\sigma/\pi \simeq 15-16$.

We hereafter refer to  $q(\beta)$ as the normalization of the diffusion coefficient such that
\beq
\label{Eq:DdE}
D(E)={q(\beta) \over \pi} \times {\ell_{\rm coh} c \over \eta_{\rm T}} \times \left({\bar{r}_{\rm L} \over 
\ell_{\rm coh}}\right)^{2-\beta} \ .
\eeq
 It is noteworthy that the normalization of the diffusion coefficient is given by $q(\beta)$ and must not be confused with the normalization of the turbulent spectrum. Both quantities appear to have similar  expressions as seen from quasi-linear theory calculations or from numerical estimates obtained in \cite{Casseetal02}. Nevertheless, they differ in a strong turbulence regime. \cite{Reville08} discussed some solutions by clearly displaying diffusion coefficient with sub-Bohm values. This issue is beyond the scope of the calculations performed here and its consideration is postponed to a future work (see also a recent work by \cite{Shalchi09}). Considering these uncertainties,  we consider $q(\beta)$ as a free parameter hereafter.

\cite{Pelletieretal06} obtained a 1D stationary $\beta = 2$ power-law solution for the non-resonant wave spectrum. We can see from the above analysis 
that the  energetic particle transport properties around the shock front depend on the possibility that non-resonant 
instability will deeply enter in the non-linear regime.  Verifying this condition leads to a diffusion coefficient at $E \ll E_{\rm CR-max}$  given 
by Eq.(\ref{Eq:D2b}),the magnetic field profile being characterised by an exponential growth over a scale $x_{\rm g}$ and a linear growth over 
a scale $x < \ell_{\rm diff}(E_{\rm CR-max})$.

This qualitative analysis confirms that the non-resonant instability contributes to the turbulence level over a wide range of parameters (once
the non-linear regime of the instability is established) and the control of the turbulence coherence length. However, the analysis presented in \cite{Pelletieretal06} 
shows however that the resonant instability at least in the domain 2 of our fiducial example above also contributes to the magnetic fluctuation spectrum. The resonant wave spectrum is
found to be harder, i.e., for a CR distribution spectrum scaling as $p^{-4}$, the 1D turbulence spectrum has an index $\beta = 1$. In this work we assume that the turbulence index is in  the range $1 \le \beta \le 2$. 
\subsection{Shock particle distribution}
\label{S:Pdif}
Before discussing the effect of turbulence evolution in the downstream region, we present the general solution for the particle distribution at the shock front in the case of spatially varying diffusion coefficients, where radiative losses are discarded. The complete calculation is presented in Appendix \ref{S:Appb}. We briefly outline our result (see Eq. \ref{Eq:gene3}) as follows. We have assumed upstream and  downstream magnetic fluctuations variation lengths $\ell_{\rm u/d}$ to  be scale (or energy) dependent (see section \ref{S:DoMF}). The slope of the stationary particle distribution (neglecting any radiative loss)  at the shock front is
\bea
\label{Eq:gene3b}
\frac{d \ln f_{S}(p)}{d \ln p} & = &-\frac{3 r}{r-1} \times \left[ \frac{D_{\rm u}(0,p) \exp\left(\int_{\rm -\ell_{\rm u}}^0 \theta_{\rm u}(x',p) dx'\right)}{u_{\rm u} 
\int_{-\ell_{\rm u}}^0\exp\left(\int_{\rm -\ell_{\rm u}}^x \theta_{\rm u}(x',p) dx'\right) dx}\right.  \nonumber \\
& & + \ \left.\frac{D_{\rm d}(0,p) \exp\left(-\int_0^{\rm \ell_{\rm d}} \theta_{\rm u}(x',p)dx'\right)} {r u_{\rm d} \int_0^{\ell_{\rm d}} \exp\left(-\int_x^{\rm \ell_{\rm d}}\theta_{\rm d}(x',p) dx'\right) dx} \right] \ .
\eea
The value of the spectrum slope  is controlled by the functions $\theta_{\rm u/d}= u_{\rm u/d}/D_{\rm u/d}-d\ln D_{\rm u/d}/dx$ (see Eq.\ref{Eq:theta}).
In the basic case where both upstream and downstream diffusion coefficients can be assumed as space independent over lengths   $\ell_{\rm u/d}$ from the shock (and to vanish beyond these distances), the above expression reduces to \citep{Ostrowski96}: 
\beq
\label{Eq:OS}
\frac{d \ln f_{S}(p)}{d \ln p}  =  - \frac{3}{r-1} \left(\frac{r}{1-\exp(-u_{\rm u}\ell_{\rm u}/D_{\rm u})}+\frac{1}{\exp(u_{\rm d} \ell_{\rm d}/D_{\rm d})-1}\right) \ .
\eeq
If the shock wave is modified by the CR back-reaction, $r$ will depend on the particle energy and the shock spectrum will not behave like power law. We note that provided functions $\theta$ are large compared to unity,  the previous relation indicates that we obtain standard power-law spectrum expected from DSA theory.

The present article investigates the effects of energy and spatial dependencies of the $\theta$ functions in both up- and downstream regions, by relying on a set of available multiwavelength data of five SNR:
Cassiopeia A, Tycho, Kepler, SN1006, and G347.3-0.5 (also known as RXJ 1713-3946.5). All of these remnants are of the case $2$ discussed in sect. \ref{S:MFr} and correspond to mildly fast shocks where both resonant and non-resonant magnetic field amplification occur.

\section{Particle acceleration in the case of a downstream advected magnetic field}
\label{S:Cmfp}
This section examines the DSA process for an efficient turbulence amplification mechanism producing a strong magnetic field in the shock precursor (see sect. \ref{S:Upa}).
In the first sect. (\ref{S:Adva}), we reconsider the calculations performed by \cite{Parizotetal06} but this time including the effect of turbulent scale compression at the shock front.
Section (\ref{S:FII}) then addresses the usually overlooked aspect of stochastic particle acceleration in the downstream flow. Finally, sect. (\ref{S:Numtest}) deals with tests involving the shock solutions obtained by \cite{Zira07} for various turbulent spectrum scalings. We then incorporate particle losses and Fermi stochastic acceleration into the Fermi cycles and proceed with different numerical experiments. We conclude with a comparison between X-ray and $\gamma$-ray filaments produced by inverse Compton up-scattering  of cosmic microwave background photons.

\subsection{Downstream diffusion regimes and maximum particle energies}
\label{S:Adva}
Downstream of the shock, the particle distribution was fully isotropised (to an order of $V/v$) and the streaming instability quenched. We insert the magnetic profiles derived in the previous section into the diffusion coefficients (see Eq.\ref{Eq:D2b}). To derive the downstream diffusion coefficients,  we need to specify properly how the transition occurs at the shock front. We only consider here the case of a strong magnetic field amplification at the shock precursor. The upstream  magnetic field being highly disordered, the magnetic compression ratio then becomes $r_{\rm B} = \sqrt{(1+2r_{\rm sub}^2)/3} \le r_{\rm sub}$ (with $r_{\rm sub} \ge 1$)\footnote{We make a distinction between the compression ratio at the sub-shock ($r_{\rm sub} \le 4$) and the total shock  compression ratio $r_{\rm tot} \ge 4$. In the case of weakly modified shocks, we have $r_{\rm tot} \simeq r_{\rm sub} \simeq r = 4 $. In the case of strongly CR modified shocks, one obtains $r_{\rm tot} > r > r_{\rm sub}$. If the sole adiabatic heating of the precursor is considered, values $r_{\rm sub} =2-3$ and $r_{\rm tot} > 10$ are possible (see e.g., \cite{Berezhko99}). If a substantial gas heating in the precursor  is produced for instance by the absorption of Alfv\`en waves, the total compression ratio cannot be much higher than 10, under ISM conditions considered above \citep{Bykov04}. In a strongly modified shock, the most energetic electrons producing the X-ray filaments have energy $\rm{E \gg E_{\rm CR min}}$ and do experience a compression ratio close to $r_{\rm tot}$. This value will be used in the following estimations. Values of $r_{\rm sub} = 2$ and  $r_{\rm tot} = 10$ are accepted in this work in the case of strongly CR modified shock.}
\beq 
\label{Eq:Bu}
B_{\rm{u}} = B_{\rm{d}} \times \left({1+2 \ r_{\rm sub}^2 \over 3}\right)^{-1/2} = {B_{\rm{d}} \over r_{\rm B}} \ .  
\eeq 
\cite{Parizotetal06} only considered this final effect. But in the meantime, the maximum turbulence
scale downstream is reduced  by a factor $r_{\rm sub}$: 
\beq
\label{Eq:Lam} 
\lambda_{\rm max-d} = {\lambda_{\rm max-u} \over r_{\rm sub}} \ .  
\eeq 
This scale compression induces an enhancement of the tangential magnetic field component and a reduction in the maximum turbulence length in the downstream region. The downstream turbulence  is then anisotropic, displaying elongated eddies in the direction parallel to the shock front \citep{Marcowithetal06} unless other non-linear processes prevail \citep{Zira08}. The coherence length of the turbulence is hereafter assumed to be a constant.

We can define the downstream diffusion coefficient according to the definition of the upstream coefficient given 
in Eq.(\ref{Eq:D2b})
\beq 
\label{Eq:ddo}
D_{\rm d} = {q(\beta) \over \pi} \times {\rho_{\rm M} \lambda_{\rm max-d} c \over 2\pi \eta_{\rm T-d}} \times
\left({\rho_{\rm Ld} \over \rho_{\rm M}}\right)^{2-\beta_{\rm d}}  \ , 
\eeq
In the remaider of the article, we only consider the case where $\beta_{\rm u} = \beta_{\rm d} = \beta$.
  
Using Eq.(\ref{Eq:D2b}) evaluated at $x=0$ as well as  Eqs. (\ref{Eq:Bu}) and (\ref{Eq:Lam}), we end up linking up- and
downstream diffusion coefficients at the shock front (where we have assumed that $\eta_{\rm T} \simeq 1$):
\beq
\label{Eq:D}
D_{\rm u} = D_{\rm d} \times r_{\rm sub} \left({r_{\rm B} \over r_{\rm sub}}\right)^{2-\beta}  = D_{\rm d} \times H(r_{\rm sub},\beta) \ , 
\eeq
Once the up- and downstream diffusion coefficients are set, the magnetic field at the shock front can be inferred following 
the same procedure as that adopted in \cite{Parizotetal06} (see the article for the detailed derivation). The balance between the electron acceleration rate and the mean synchrotron loss rate fixes the maximum electron energy to $t_{\rm{acc}}(E_{\rm{e-max}}) = \langle t_{\rm{syn}}(E_{\rm{e-max}})\rangle $. 
The synchrotron loss timescale is obtained from Eq.(17) of \cite{Parizotetal06} using the mean square magnetic field experienced by relativistic electrons during one Fermi cycle: 
\beq
\label{Eq:mB}
\langle B^2 \rangle = B^2_{\rm{d}} \times \left({H(\beta)/r_{\rm
    B}^2+r_{\rm tot} \over H(\beta) + r_{\rm tot}}\right) \ .  
\eeq 
Following DSA standard theory the acceleration rate is  
\beq 
\label{Eq:tauacc}
t_{\rm{acc}}(E) = {3 r^2 \over r-1} \ {D_d(E) \over V_{\rm{sh}}^2}
\times \left[{H(r,\beta)\over r(E)} +1\right] \ .   
\eeq  
Basic analytical relations can be derived when Bohm diffusion regime conditions prevail. In that case, electron and  proton accelerations  are no longer related because the diffusion coefficient
no longer depends on $\lambda_{\rm max}$ anymore\footnote{excepted at the highest energies.}. Equation (30) in \cite{Parizotetal06} can
be used to derive the downstream magnetic field amplitude and an estimate of the synchrotron photon energy cut-off
\beq   
E_{\rm \gamma -cut} \simeq [0.875 \ {\rm \rm{keV}}] \times \frac{V_{\rm sh,4}^2}{\bar{q} \bar{y}(r_{\rm tot})
(1 + H/r_{\rm tot} r^2_{\rm B})} \ ,
\label{Eq:Ecuthyp}
\eeq  
where we note that $\bar{y}(r_{\rm tot}) = 3 \bar{r}_{\rm tot}^2/(r_{\rm tot}-1)$, 
$\bar{r}_{\rm tot}=r_{\rm tot}/4$, and $\bar{q} = q(\beta=1)/16$. The maximum electron
energy is found to be approximately 10 TeV in our SNR sample, a value close to the maximum CR energy. To derive such result, we assumed that the compression ratio $r$ at $E=E_{\rm e-max}$ is approximatively  $\sim r_{\rm tot}$.

\begin{table*}[t]
\begin{center}
\begin{tabular}{|c||c|c|c|}
\hline  Supernova remnant &   $B_{\rm d} \ (\mu G)$ &  $\displaystyle \frac{E_{\rm\gamma-cut}}{E_{\rm\gamma-cut,obs}}$Ê &  $B_{\rm d-FII}\ (mG)$ \\ \hline
Cas A & $558$ & $0.2$ & $2.7$ \\ \hline Kepler & $433$ &  $0.3$ & $2.3$ \\ \hline Tycho &
$586$ & $0.7$ & $1.5$ \\ \hline SN 1006 & $170$ & $0.07$ & $0.56$ \\ \hline G347.3-0.5 &
$131$ & $0.05 $ &  $2.1$ \\ \hline
\end{tabular}
\end{center}
\caption{Inferred values of the downstream magnetic field amplitude and synchrotron
  photon cut-off energy in the case of an {\it advection-dominated} rim where Bohm diffusion regime prevails ($\beta=1$ and $q(\beta=1)=\sigma$). The magnetic field values were calculated by assuming  $r_{\rm tot}=r_{\rm sub} = 4$. }
\label{T:tab1}  
\end{table*}
In Table (\ref{T:tab1}), we list the inferred values of the downstream magnetic
field in the context of an advection dominated X-ray rim, where a Bohm-type
turbulence is occurring. We have also provide the theoretical values of
$E_{\rm \gamma -cut}$ required  to verify $t_{\rm acc}(E_{\rm e-max})=\langle 
t_{\rm syn}(E_{\rm e-max}) \rangle$. The similar to those in table 1 of
\cite{Parizotetal06}, except for SN1006 where we used the value of shock velocity ($4900 km/s$) given in \cite{Aceroetal07}. The results presented in this table were performed using a diffusion coefficient normalization $q(\beta=1)=\sigma$ corresponding to predictions by the quasi-linear theory.

Based on the aforementioned assumptions,  it appears that older SNRs ($T_{\rm SNR}> 1000 \ \rm{yr}$) should have a synchrotron cut-off energy that is much lower than the observed value. However, as for instance in the case of  SN1006, the cut-off frequency depends on the observed region of the SNR and 3 keV is probably an upper limit. On the other hand, young SNRs ($T_{\rm SNR} < 500 yr$) exhibit, in the same context, strong magnetic fields and synchrotron energies cut-off close to the cut-off deduced from the observations. The effect is even stronger in the case of modified shocks. \cite{Parizotetal06} noted that the Bohm regime does not allow the DSA theory to reproduce accurately the X-ray filaments unless the diffusion coefficient normalization is replaced by a factor $k_0$ of a few. This is confirmed by the close agreement between the two cut-off energies obtained for the young SNR.

Several uncertainties may shift the value of the cut-off frequency from the extrapolation using the radio data. \cite{Zira07} pointed out that the electron particle distribution can be cut off in a smoother way than by a pure exponential cut-off. In that case, the true cut-off frequency is shifted towards higher energies. In the meantime, the observed synchrotron cut-off used previously is probably to be an upper limit because of the back-reaction of CR on the shock structure producing a curved shape of the spectrum. It seems justified to develop a detailed non-linear calculation to improve the estimate of the discrepancy between these solutions with a simple exponential cut-off. This aspect should also be an important issue for the next hard X-ray satellites generation such as nuStar or Next. We postpone its investigation to future work.

To summarise, we can say that scale compression has a very limited impact on the above calculation and that the results derived in \cite{Parizotetal06} are quite robust. 

\subsection{Considering downstream stochastic Fermi acceleration}
\label{S:FII}
The downstream magnetic field amplitudes derived  in sect. \ref{S:Cmfp} are lower limits, while the observed filament sizes are just upper limits because of the lack of resolution of X-ray instruments. If the downstream magnetic field reaches values close to mGauss and does not relax rapidly, then at some stage the Alfv\'en velocity will be of the order of the downstream fluid velocity. In that case, stochastic Fermi acceleration can no longer be neglected.
Electrons will interact with turbulence modes generated by the resonant streaming instability  since  non-resonant modes are right-handed polarized and thus cannot interact with electrons. We included in our numerical calculations the so-called Fermi second-order process (in addition to the usual first-order acceleration) combined with energy losses, namely synchrotron losses for the electrons. We implicitly assume in the remainder of the paper that an efficient redistribution of forward and backward waves operates by means of non-linear interaction with magneto-sonic waves \citep{Pelletieretal06}. In that case, forward and backward modes transmitted downstream are in balance \citep{Vainio99}. This assumption enables us to estimate the magnetic field amplitude when dominant stochastic Fermi acceleration occurs. Issues dealing with imbalanced magnetic turbulence are beyond the scope of this paper and will be investigated in future work.\\
The  acceleration timescale characterising the stochastic Fermi process for a relativistic particle can be written as
\beq
\label{Eq:TF2}
t_{\rm acc, FII} \simeq {9 D(E) \over V_{\rm A,d}^2} \ .
\eeq
The conditions in which a stochastic acceleration less efficient than the usual shock acceleration can be transposed into a condition on the downstream magnetic field  by writing  $t_{\rm acc, FII} \le t_{\rm acc, FI}$. Using Eqs. (\ref{Eq:tauacc}) and (\ref{Eq:TF2}), one can readily find that
\beq
B_{\rm d-FII} \le [714 {\rm \ \mu Gauss}] \times {n_{\infty,-1}^{1/2} V_{\rm sh,3} \bar{r}_{\rm tot}^{1/2} \over \bar{y}(r)^{1/2} (H(r_{\rm tot},\beta)
/r_{\rm tot}+1)^{1/2}} \ ,
\eeq
In this expression, we exceptionally use a shock velocity expressed in units of $10^3\ \rm{km/s}$ and ISM density in units of $0.1 \ \rm{cm^{-3}}$.

In the case of young SNRs propagating into a standard ISM medium with typical hydrogen densities $\sim 10^{-1} \ {\rm cm}^{-3}$, the previous
limit leads to magnetic field strengths $\sim 1-2$ mGauss for a typical shock velocity of the order of $5\times 10^3$ km/s.  This is confirmed by the values of the limited magnetic field strengths given tab. \ref{T:tab1} for each SNR. The surrounding gas density in most cases provides only a crude estimate or is derived from averaged values over the entire remnants. We have used for Cas A: $n_{\rm \infty} = 1$ \citep{Berezhkoetal03b}, Kepler: $n_{\rm \infty} = 0.7$ \citep{Aharonianetal08},  Tycho: $n_{\rm \infty} = 0.4$ \citep{Hughes00}, SN1006: $n_{\rm \infty} = 0.05$  (SE rims see \cite{Aceroetal07}), G347.3-0.5: $n_{\rm \infty} = 1$ (poorly constrained see \cite{Aharonianetal06}).

The Fermi stochastic acceleration process produces an energy gain in the downstream medium and a hardening of the particle distribution at the shock front (see Eq. 15 in \cite{Marcowithetal06} and the
simulations in sect. \ref{S:SFII}). As particles are continuously reaccelerated downstream, they are expected to produce larger X-ray filaments. Both effects seem clearly incompatible with the available data. The magnetic field fluctuations in resonance with electrons are then expected to saturate at the shock front with magnetic field amplitude $\ll B_{\rm d-FII}$, which is below the value for equipartition with thermal pressure of the flow. 

\subsection{Numerical experiments}
\label{S:Numtest}
The SDE method presented in Appendix \ref{S:Appc} does not account for the back-reaction of CR 
over the fluid flow. This would require a special smoothing and a difficult treatment of the CR pressure $P_{\rm CR}$. 
The latter calculated from the particle distribution $f(p,r)$ at each grid point would produce unphysical 
fluctuations that develop with time. Several numerical works have started to included wave generation effects in CR modified shock hydrodynamics \citep{Vladimirovetal06, Kang07, Vladimirovetal08}. Some semi-analytical works has also started to investigate
the effect of the wave precursor heating on the CR back-reaction process \citep{Capriolietal08a}. Both approaches seem to reach a similar conclusion: the heating of the precursor by the wave damping reduces the gas compressibility and thus reduces the shock compression \citep{Bykov04}. Stationary solutions are found to be rather close to the test particle case. Calculations performed in the test particle framework using SDEs can then reproduce the main properties of the particle acceleration process.
SDE have several advantages: they are simple to implement and rather simple to couple with  MHD equations.  SDE schemes enable a fast and large investigation of the parameter space of the DSA mechanism. For instance, the inclusion of Fermi stochastic 
acceleration is rather simple in both the SDE scheme and in the use of various spatial diffusion coefficient regimes.
Our results can, for instance, be used as limiting tests for future non-linear simulations.

\subsubsection{Synchrotron spectrum solutions}
\begin{figure}
  \resizebox{\hsize}{!}{\includegraphics{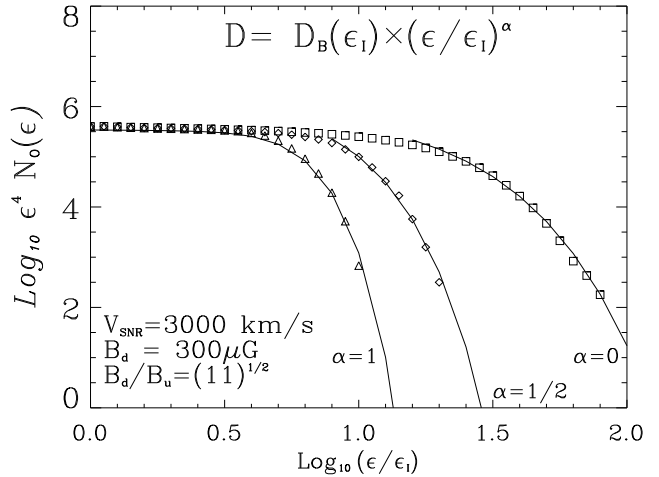}}
   \caption{Shock front energy spectra of relativistic electrons provided by multi-scale
     simulations where the MHD part of the simulation mimics the behavior
     of a SNR blast wave.}
   \label{Fig30}
\end{figure}

We first validate the aforementioned numerical scheme by achieving calculations in different configurations,
such as reproducing the analytical results of \cite{Zira07}. In this work,
the authors define the relativistic electron energy spectra
at the shock front in the presence of a discontinuous magnetic field  (the
discontinuity is located at the shock). We performed several
SDE-MHD simulations where constant upstream and downstream magnetic
fields  prevail ($B_d/B_u=r_B=\sqrt{11}$, $r_{\rm sub}$ is set to 4) and where the 
shock velocity of the flow is set to 3000 km/s. The various presented simulations differ only 
in terms of their implemented spatial diffusion coefficients, where $D=D_{\rm Bohm}(E_{\rm inf}) (E/E_{\rm inf})^{\alpha_{\rm D}}$ 
(the particles are injected at energy $E_{\rm inj}=5$ TeV).  \cite{Zira07} provided the shape of the electron energy spectra at the
shock front beyond the energy cut-off $E_{\rm e-max}$ induced by synchrotron losses, namely $N(E) \propto \exp(-(E/E_{\rm e-max})^{1+\alpha_{\rm D}})$. 
Figure \ref{Fig30} displays three simulations with $\alpha_{\rm D}=1$ (Bohm diffusion), $\alpha_{\rm D}=1/2$ (Kraichnan turbulence), and $\alpha_{\rm D}=0$ (constant coefficient). The result of the numerical calculations are displayed using items, while analytical solutions of \cite{Zira07} are displayed using solid lines. 
In the figure the following parameters have been used: velocity of the downstream fluid 3000 km/s, compression factor $r_{\rm sub}=4$) and uniform upstream and downstream magnetic field are set ($r_B=\sqrt{11}$.  We have set various diffusion regime ($D\propto E^{\alpha_{\rm D}}$) while using our new numerical SDE scheme described in appendix \ref{S:Appc}. The agreement between numerical calculations and analytical profile is good and proves that the skew SDE numerical scheme is valid for all kinds of diffusion regimes and can handle magnetic discontinuities properly (see sect. (\ref{S:SDE2D}) for further details).
\subsubsection{Shock particle distribution and second order Fermi process}
\label{S:SFII}
\begin{figure}
  \resizebox{\hsize}{!}{\includegraphics{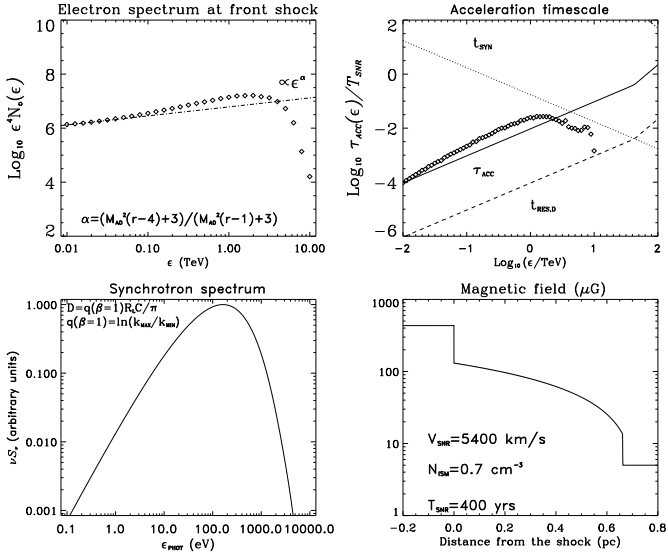}}
   \caption{Energy spectrum of relativistic electrons at the shock front given by MHD-SDE
     simulations in the conditions of the Kepler SNR.}
   \label{F:F31}
\end{figure}

\begin{figure}
  \resizebox{\hsize}{!}{\includegraphics{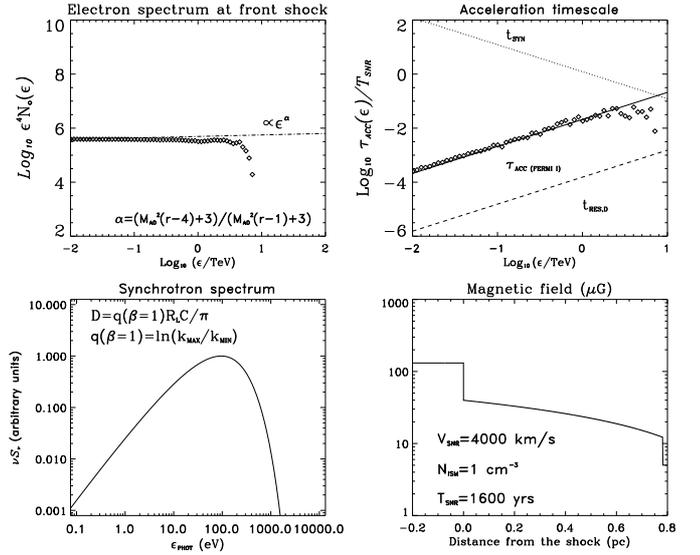}}
   \caption{Same plots than in Fig.\ref{F:F31} but in the G347-0.5 SNR (velocity of the shock is $4\times 10^3 km/s$ and compression ratio $r_{sub}=4$). The density of ISM is  $1 \rm{cm^{-3}}$. The Bohm regime for 
     the diffusion coefficient has been assumed with $q(\beta=1)=15$. The simulation has been performed until time $t=1600$ years.}   \label{F:F32}
\end{figure}

Figure \ref{F:F31} and \ref{F:F32} show the shock particle distribution and synchrotron spectra for the parameters 
corresponding to the conditions that prevail in the Kepler and G347.3-0.5 SNRs, respectively. In the case of the Kepler SNR, we use the parameters
$V_{\rm sh}= 5.4\times 10^3$ km/s, $B_{\rm d} = 433 \ \mu G$ and, $\beta =1$. Upstream 
density is $0.7 \rm{cm}^{-3}$ (\cite{Berezhkoetal06} estimated the density to be $n_{\infty} \le 0.7 
\rm{cm}^{-3}$). In the case of G347-0.5, we set parameters to be $V_{\rm sh}= 4000$ km/s, $B_{\rm d} = 131 \ \mu G$, $\beta =1$. The averaged upstream density is $1\  \rm{cm}^{-3}$ \citep{Aharonianetal06}. In both cases, the magnetic profiles used in the simulations are also presented.  The dashed-line shows the stationary solution found in \cite{Marcowithetal06}, which includes particle re-acceleration in the Fermi cycle. In the upper right panel, the acceleration (with the sole regular Fermi acceleration), and both the diffusive and downstream residence timescales are displayed. Diamonds are obtained using a numerical calculation of the acceleration timescale. The slight excess is produced by the stochastic Fermi acceleration process.  We also display the synchrotron spectrum and the magnetic profile around the shock front at $t=400 yr$.

The maximum CR energy (and the aspect ratio $k_{\rm max}/k_{\rm min}$) corresponds to the maximum CR energy limited by either particle escapes in the upstream medium or the SNR age limit.  At $E_{\rm CR-max}$, the maximal upstream diffusion coefficient allowed by the escaping limit is:
\beq
\label{Eq:DRV}
D(E_{\rm CR-max})= \chi \times R_{\rm sh} \ V_{\rm sh} \ .
\eeq 
The factor $\chi$ is usually not accuratly defined. An accurate determination of this parameter would require to be performed by non-linear
simulations of DSA that include the effect of the turbulence generation back-reaction on the flow. A  fraction of few tenth of percent of the SNR radius is usually assumed in theoretical calculations and seems to be reasonable \citep{Berezhko96, 
Capriolietal08b}. The normalization $\bar{\chi}=\chi/0.3$ is then acceptable in this text.

It can be seen from Figs.\ref{F:F31} and \ref{F:F32} that stochastic acceleration slightly modifies the shock particle spectrum in the case of the Kepler SNR. The synchrotron losses create a bump close to the maximum
electron energies. In the Kepler remnant, the synchrotron cut-off is found to be around 0.2 keV (see Fig.\ref{F:F31}), while in the case of G347.3-0.5 it is 
around 0.5 keV (see Fig.\ref{F:F32}). We verified that lowering the normalization factor $q(\beta)$ of the diffusion coefficient from $16$ to $3$ produces a cut-off around 1 keV (Kepler) and 2.5 keV (G347.3-0.5), namely that a higher cut-off would require a lower $q(\beta)$ (see Eq. \ref{Eq:Ecuthyp}). The density around G347.3-0.5 is badly constrained and $n_{\infty} < 1 \rm{cm^{-3}}$ would lead to similar effects. We note that the above simulations maximize the incidence of the stochastic acceleration because we assumed that the resonant field dominates the total field in the downstream medium (see Eq. \ref{Eq:BRBNR}).

To conclude it clearly appears that the downstream Alfvenic Mach number $V_{\rm d}/V_{\rm A,d}$ cannot be much less than unity otherwise: (1) the X-ray filament would be too large with respect to the observed widths (see next section),  (2) the X-ray cut-off frequency would be far larger than $E_{\rm \gamma,cut}$ (see Fig.\ref{F:F31}), and (3) the radio spectrum would be harder than $\nu^{-0.5}$ (see Fig.\ref{F:F31}). Generally speaking, the maximum downstream resonant magnetic field cannot be much stronger than a few mGauss downstream of the shock front, otherwise regular acceleration process would be dominated by stochastic Fermi acceleration. This places an important constraint on the combined value of the magnetic field and the local ISM density as well as the respective contribution of the resonant and the non-resonant instability to the total magnetic field at the shock front.


\subsubsection{Comparisons between X- and $\gamma$-ray filaments}
We end this section by a detailed comparison between X- and $\gamma$-ray filaments produced by the relativistic electrons. The inclusion of neutral pion decay caused by the hadronic
interaction with the interstellar fluid or with the shocked matter would require a complete modelling of both the hadron spectrum and the ISM density profile around the SNR. This study is postponed to future work. 

\begin{figure}[t]
  \resizebox{\hsize}{!}{\includegraphics{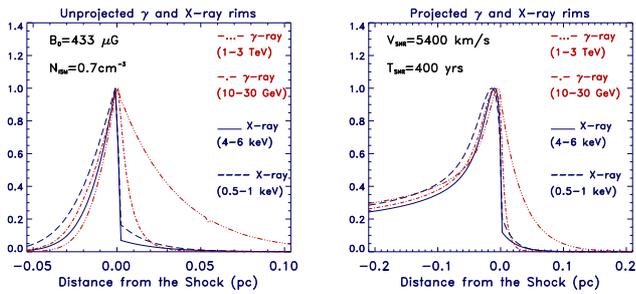}}
   \caption{The unprojected and projected X-ray and $\gamma$-ray rims in the conditions of the Kepler SNR (same physical conditions than Fig.\ref{F:F31}). For clarity, both X- and $\gamma$-ray rims have been normalized to one.}
   \label{F:F33}
\end{figure}

\begin{figure}[t]
  \resizebox{\hsize}{!}{\includegraphics{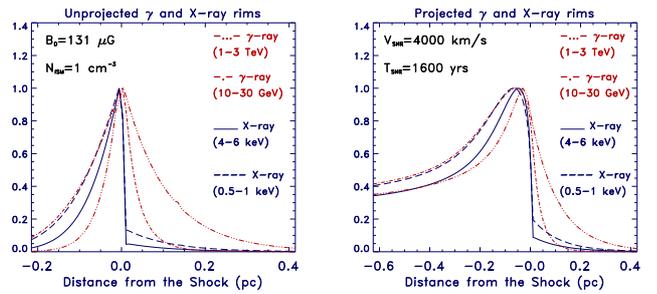}}
   \caption{The unprojected and projected X-ray and $\gamma$-ray rims in the conditions of the G347-0.5 SNR (same physical condition than in Fig.\ref{F:F32}). For clarity both X- and $\gamma$-ray rims have been normalised to one. }
   \label{F:F34}
\end{figure}

In our calculations, the leptonic $\gamma$-ray emission was integrated into two characteristic wavebands 10-30 GeV and 1-3 TeV using the standard expression of the isotropic inverse Compton emissivity \citep{Blumenthal70}. The rims are produced by the scattering off the cosmic microwave photons by relativistic electrons. They are displayed in Figs.\ref{F:F33} and\ref{F:F34}, where they were obtained with parameters adapted to the dynamics of the Kepler and the G347-0.5 SNR, respectively. We also displayed  two X-ray wavebands ( $4-6$ keV and $0.5-1$ keV, even if this later wave band is usually dominated by the thermal emission). In each case, both projected and deprojected filaments are reproduced. 
The relative normalization between X-ray and $\gamma$-ray filaments depends mostly on the intensity of the magnetic field; for the same particle energy domain, it is found to scale as $B^2$ as expected. The width of the $\gamma$-ray TeV rim is usually the greatest because an important fraction of the IC radiation is produced upstream. The $10-30$ GeV $\gamma$-rays are produced closer to the shock upstream than $1-3$ TeV $\gamma$-rays. In the downstream region, the highest energetic electrons are confined closer to the shock because of their shorter radiative loss timescales. The projected rims indicate that only a slight difference exists between the position of the peak of the gamma and X-ray emission.  As the size of the $\gamma$-ray rims is not much larger than the X-ray filaments, it seems impossible for any $\gamma$-ray instrument to separate both components. This will also be the case for future instruments such as CTA unless the filaments are very large (see the case of Vela Junior discussed in \cite{Bambaetal05a}). 

\section{Diffusive shock acceleration in the case of downstream spatially relaxing turbulence}
\label{S:DoMF}
We now consider a scenario where the downstream magnetic field fluctuations vary over a length-scale much shorter than the SNR shock radius $R_{\rm SN}$. This scale noted $\ell_{\rm d}$ can depend on the wave number $k$ of the fluctuations. The damping of the turbulence in the downstream medium and its compression at the shock front can modify the particle mean residence time and the relativistic particle return probability to the shock. Hence, this magnetic relaxation is expected to modify the efficiency of the diffusive acceleration process itself.

Equation (\ref{Eq:OS}) shows that the particle energy spectrum at the shock front remains a power law, provided that the quantities (at a given energy $E$) $z_{\rm u/d}(E)= u_{\rm u/d}\ell_{\rm u/d}/D_{\rm u/d}$ are large compared to unity. For $z_{\rm u/d}(E) \le 1$,  the particle distribution will be strongly softened  and the acceleration timescale will shorten dramatically, the latter being dominated by the particles experiencing the shortest residence time. A softening effect induced by the upstream losses is only expected at the highest energy close to $E_{\rm CR-max}$, namely as $z_{\rm u} \rightarrow 0$ \footnote{Again, a correct way to handle this effect is to account properly for the particle back-reaction on the flow.}. The diffusive length of particles with energy lower than $E_{\rm CR-max}$ is always smaller than the variation scale of the magnetic fluctuations $\ell_{\rm u}$ (controlled by the highest energy), hence we have $z_{\rm u} (E < E_{\rm CR-max}) \gg 1$, leading to a vanishing exponential factor in the above solution. In contrast, the softening effect downstream can be significant at energies much lower than $E_{\rm CR-max}$ as $\ell_{\rm d}$ can be highly scale (and thus energy) dependent. This is precisely the main topic of 
this section, namely  trying to identify the parameter space that allows the Fermi acceleration process to be efficient in the context of  a relaxing downstream turbulence.
 
 Hereafter the downstream relaxation length $\ell_{\rm d}$ is considered to be energy dependent and we normalize it with respect to the maximum CR energy; 
$E_{\rm CR-max}$: 
\beq
\label{Eq:Elld}
\ell_{\rm d}(E) = \ell_{\rm d,M} \times \left({E \over E_{\rm CR-max}}\right)^{\delta_{\rm d}} = \ell_{\rm d,M} \times \left({k_{\rm min} \over k}\right)^{\delta_{\rm d}} .  
\eeq  
The scale $\ell_{\rm d,M}$ is the relaxation scale at the maximum particle energy \footnote{All quantities with an index M are to be taken at the maximum particle energy.}. We reiterate recall that the relationship between energy particle and wave vector originates from the condition that a given particle should resonate with a turbulence mode,  $kr_L\sim 1$. We first investigate the magnetic field profiles in the downstream medium  resulting from various relaxation processes (Sect. \ref{S:MFd}). In Sect., \ref{S:Arc} the efficiency of the DSA with respect to the turbulence properties (turbulence index, relaxation index) is discussed, by considering in particular the effect of the downstream magnetic field amplitude. Various numerical experiments, presented in Section \ref{S:Numr} illustrate the effect of the magnetic field spatial variation in the particle dynamics and the associated X- and $\gamma$-ray rims.

\subsection{Downstream magnetic field relaxation}
\label{S:MFd}
This work considers various turbulent magnetic field damped profiles: the case of an energy-dependent Heaviside profile, the profile produced by a non-linear Kolmogorov-type damping \citep{Ptuskin03}, and the profile produced by the Alfv\'en or fast magnetosonic
cascades \citep{Pohletal05}. We also briefly discuss the case of a turbulent dynamo action downstream \citep{Pelletieretal06}. In this section, unless specified otherwise $\delta_{\rm d} \ge 0$ is implicitly assumed.

\subsubsection{Heaviside profiles}
\label{S:Hp}
Heaviside-type magnetic relaxation accounts for  an idealized approach to turbulence relaxation, where a given turbulence mode is assumed to be uniform out a distance $\ell_{\rm d}(k)$ from the shock and to vanish beyond that distance. This relaxation model is probably unphysical but enables us to reproduce the basic features of the turbulence relaxation effects upon particle acceleration.  Assuming this profile, we write the magnetic energy  turbulence spectrum as (the downstream medium is defined by $x>0$)
\beq
\label{Eq:pbdo}
W(k,x)_{\rm d} =W(x = 0^+,k)\Pi(\ell_{\rm d}(k)-x) + W_{\infty}\Pi(x-\ell_{\rm d}(k)) \ ,
\eeq   
where $\Pi$ functions are Heaviside functions and $x$ is the distance from the shock front. The magnetic energy density far downstream is $W_{\rm\infty}$. The normalization 
of the turbulent spectrum $W(k)=W_0 \bar{k}^{-\beta}$ is related to the magnetic field at the shock front 
by means of  $W_0 = B^2(x=0^+)/4\pi \sigma k_{\rm min}$,  where $k_{\rm min} = 2\pi \lambda_{\rm max,d}^{-1}$, $\bar{k}=k \lambda_{\rm max,d}$, 
and again $\sigma(\beta=1)=\ln(k_{\rm max} /k_{\rm min})$ and $\sigma(\beta>1)\simeq 1/(\beta-1)$.

The Heaviside profile, despite it crudely approximating the variation in the magnetic 
energy density downstream, permits us to derive a basic spatial profile of the total magnetic field
given by 
\beq
\label{Eq:DeltaB}
{\delta B^2(x) \over 4\pi} = \int_{\rm k_{\rm min}}^{k_{\rm max}} W(k,x) dk \ ,
\eeq
which in the case of Bohm turbulence leads to ($\ell_{min}$ is defined as $\ell_{\rm d,M}\times (k_{min}/k_{max})^{\delta_d}$)
\begin{eqnarray}
\label{Eq:DeltaB2}
0<x\leq \ell_{\rm min} \ \ &:& \ {\delta B^2(x) \over 4\pi} = {\delta B^2(0^+) \over 4\pi} + {\delta B^2_{\infty} \over 4\pi}\nonumber \\
\ell_{\rm min}\leq x\leq \ell_{\rm d,M} \ &:& \ {\delta B^2(x) \over 4\pi} = {\delta B^2(0^+) \over 4\pi}\frac{\ln\left(\ell_{\rm d,M}/x\right)}{\delta_d\ln(k_{\rm max} /k_{\rm min})} +{\delta B^2_{\infty} \over 4\pi} \nonumber\\
\ell_{\rm d,M} \leq x \ \ &:& \  {\delta B^2(x) \over 4\pi} = {\delta B^2_{\infty} \over 4\pi} 
\end{eqnarray}
At any given downstream location $\ell_{\rm d,M}\geq x\geq\ell_{\rm min}$ , the maximum non-vanishing turbulence wave number  is $k_{\rm max}(x)= k_{\rm min} (\ell_{\rm d,M}/x)^{1/\delta_{\rm d}}$. Beyond $\ell_{\rm d,M}$, all turbulent modes vanish giving a total magnetic field $B_{\rm\infty}$ close to the ISM magnetic field value.
 The spatial variation in the magnetic field for any other diffusion regime is more complex, as  it scales as $1-(\ell_{\rm d,M}/x)^{(1-\beta)/\delta_{\rm d}}$ for $x \ge \ell_{\rm min}$. The total magnetic field is required to calculate the synchrotron losses properly, in addition to  the normalisation entering the particle Larmor radius and the local Alfv\`en velocity. Once the total magnetic field is known, we can calculate, for every relativistic particle of energy $E$, is the fraction of the total magnetic field that can resonate with this particle, namely by integrating all turbulence modes verifying $1/\bar{r}_L(E) \leq k \leq k_{\rm max}(x)$. This is achieved by computing the function $b$ defined in Eq.(\ref{Eq:func_b}). If magnetic turbulence relaxation follows a Heaviside prescription then one obtains
\begin{align}
&b(0<x\leq\ell_{\rm min},E)\simeq  {\ell_{\rm coh}\over \beta}\left({\bar{r}_L(E) \over \ell_{\rm coh}}\right)^{\beta} \\
&b(x\geq\ell_{\rm min},E)={\ell_{\rm coh}\over \beta}\left\{\left({\bar{r}_L(E) \over\ell_{\rm coh}}\right)^{\beta} - \left({\bar{r}_L(E_{\rm CR-max}) \over\ell_{\rm coh}}\right)^{\beta}\left({x \over \ell_{\rm d,M}}\right)^{\beta/\delta_d} \right\} \ .\nonumber
\end{align}
Once both the total magnetic field and function $b$ are known, it is easy to compute in our simulations both the spatial and energy diffusion coefficients of every test particle, which are mandatory to determine the particle motion and stationary particle distribution solutions in Eqs. (\ref{Eq:gene3b})
and (\ref{Eq:gene2}). The procedure is repeated in the same way for any magnetic profile.

\subsubsection{Non-linear Kolmogorov damping}
\label{S:Kol}
In models of incompressible MHD turbulence described by the Kolmogorov
energy cascade towards large wave numbers, the non-linear damping
kernel scales as $k^{5/2} W(k)^{3/2}$. Following \cite{Ptuskin03}, this
kernel can be simplified while still respecting  the spatial relaxation
profile. We have
\beq
\label{Eq:Gnl}
\Gamma_{\rm NL}(k,x) \simeq \Gamma_0 \times k^{3/2} W(k,x)^{1/2} \ , 
\eeq
where $\Gamma_0 \simeq 5\times10^{-2} \times V_{\rm a,d}/(B_{\rm d}^2/4\pi)^{1/2}$. 
Here we consider the cascade to be initiated behind the shock and use the local total magnetic field and 
Alfv\'en velocity. \\
In the shock rest-frame, the turbulence relaxation downstream (for $x > 0$) is described
by a stationary equation 
\beq
\label{Eq:PDE}
{V_{\rm sh} \over r_{\rm tot}} \times {\partial W(k,x) \over \partial x} =
-2\Gamma_{\rm NL}(k,x) W(k,x) \ , 
\eeq 
and a boundary solution  $W(k,x=0^+)= W_0 \times \bar{k}^{-\beta}$.
The solution of Eq.(\ref{Eq:PDE}) is
\beq
\label{Eq:Rnl}
W(k,x) = {W(k,x=0^+) \over \left(1+\bar{k}^{(3-\beta)/2} \displaystyle\frac{x}{x_0}\right)^2}
\ .  \eeq 
An estimate of the scale $x_0$ is (see \cite{Pohletal05})
\beq
\label{Eq:x0K}
x_{\rm 0-K} \simeq [300 \times \lambda_{\rm max,d}] \times {V_{\rm sh,4} n_{\infty}^{1/2} \bar{r}_{\rm sub}^{1/2}
 \bar{\sigma}^{1/2} \over  \bar{r}_{\rm tot}}  \times B_{\rm d,-4}^{-1}   \ .
\eeq
We used $\bar{\sigma}=\sigma/16$ and the shock velocity $V_{\rm sh,4}$ is expressed in units of $10^4 \ \rm{km/s}$. 
The downstream maximum turbulence scale $\lambda_{\rm max,d}$ can be connected to the maximum Larmor radius of CRs upstream by means of Eq.(\ref{Eq:Lam}).  
Reduced rigidity at maximal energy $E_{\rm CR-max}$ is such that $\rho_{\rm Lu} \simeq \rho_{\rm M}$ as the diffusion coefficient rapidly increases as $E^2$ beyond $E_{\rm CR-max}$. Both conditions set the maximum upstream turbulence scale $\lambda_{\rm max,u}$ and the maximum CR energy $E_{\rm CR-max}$. We find that $\lambda_{\rm max ,d} \simeq 5.2 \ r_{\rm L-max,u}/\bar{r}_{\rm sub} \bar{\rho}_{\rm M}$, where $\bar{\rho}_{\rm M} = \rho_{\rm M}/0.3$.

The relaxation scale is $\ell_{\rm d}(E) = \ell_{\rm d,M} \times (E/E_{\rm CR-max})^{(3-\beta)/2}$. The factor $\ell_{\rm d,M}$ is defined as the length 
over which turbulence level has decreased by $1/e$ compared to its value at the shock front, i.e., 
$\ell_{\rm d,M} =(\sqrt{e}-1) x_0$. The spatial dependence of the total magnetic field and function $b$ were calculated using Eqs. (\ref{Eq:DeltaB}) and (\ref{Eq:func_b}). These expressions, which are quite lengthy especially for the $b$ function in Eq.(\ref{Eq:nus}), were implemented into the code but are not explicitly given here.

\subsubsection{Exponential profiles}
\label{S:Exp}
When turbulence damping rate does not depend on space but remains dependent on wave number ($\Gamma=\Gamma(k)$), the relaxation of the downstream magnetic field follows an 
exponential cut-off on a scale length $\ell_{\rm d}(k) = r_{\rm tot} \Gamma(k)/V_{\rm sh}$. The turbulent magnetic energy spectrum is then
\beq
W(k,x) = W(k,0^+) \times \exp\left(-{x \over \ell_{\rm d}(k)}\right) \ .
\eeq 
The Alfv\'en and Magnetosonic waves cascades considered by \cite{Pohletal05} 
follow this scaling, the corresponding damping rates and expression for $x_{\rm 0}$ can easily be obtained from their Eqs. (8) and (11) respectively. 
Considering the Alfvenic cascade, we obtain
\beq
\label{Eq:x0A}
x_{\rm 0-A} \simeq \left[\bar{\rho}_{\rm M}^{1/2} \times \lambda_{\rm max,d}\right] \times {V_{\rm sh,4}  n_{\infty}^{1/2}  \bar{r}_{\rm sub}^{1/2}
  \over  \bar{r}_{\rm tot}}  \times B_{\rm d,-4}^{-1}   \ .
\eeq
The coherence scale of the downstream turbulence is $\ell_{\rm coh} = \lambda_{\rm max,d} \ \rho_{\rm M}/ 2\pi$.
The fast magnetosonic cascade leads to a similar expression except that the wave phase velocity can be approximated as $V_{\rm FM,d} = (V_{\rm A,d}^2+c_{\rm s,d}^2)^{1/2}$, $c_{\rm s,d}$ being the sound velocity behind the shock front.We note that the above expression for the Alfv\'en cascade results from the combinaition of the critical balance and the anisotropy obtained in the Goldreich-Sridhar phenomenology of strong turbulence \citep{Goldreich95}. 
 Again, the expressions of the total magnetic field and resonant field are rather lengthy and are not shown here. We note that
Eqs. (\ref{Eq:x0K}) and (\ref{Eq:x0A}) show that the Kolmogorov damping leads to a slower cascade and thus to longer relaxation scales than exponential damping. 

\subsubsection{Turbulent dynamo downstream}
\label{S:TDD}
\cite{Pelletieretal06} (see also \cite{Zira08}) discussed the action of a turbulent dynamo in the downstream medium that would lead to additional amplification of the magnetic field. The magnetic field is expected to saturate at values close to  equipartition with the dynamic gas pressure. The dynamo action is driven by the non-vanishing helicity of the non-resonant turbulent modes. 

The corresponding scale of magnetic field variation is given by the ratio of the magnetic turbulent diffusivity $\nu_{\rm t}$ to the dynamo amplification coefficient $\alpha_{\rm D}$.
The two coefficients can be expressed as \citep{Pelletieretal06}
\beq
\alpha_{\rm D} \simeq {2 c \over 3 \pi} \times \left({\bar{V}_{\rm a} \over V_{\rm sh}}\right)^2 \times \ln\left(r_{\rm L}\left(E_{\rm CR-Max}\right)/r_{\rm L}\left(E_{\rm CR-min}\right)\right) \ , 
\eeq
and 
\beq
\nu_{\rm t} \simeq {2 c \lambda_{\rm max,d} \over 3 \pi^2} \times \left({\bar{V}_{\rm a} \over V_{\rm sh}}\right)^2  \ ,
\eeq
where $E_{\rm CR-min}$ represents the lowest resonant energy. The amplification scale is then $\ell_{\rm ampl} \sim \lambda_{\rm max,d}/(\pi\phi)$. Turbulence modes of wavelength longer than $\ell_{\rm ampl}$ grow and saturate close to equipartition. Other turbulence modes are expected to dampen rapidly (over a few plasma skin depths)
because the non-resonant waves are not normal modes of the plasma, as already stated in sect. \ref{S:NRmod}.

\subsection{Particle acceleration in a relaxed-compressed turbulence}
\label{S:Arc}
In the next few paragraphs,  we present some useful analytical estimations for the analysis of the numerical
simulations presented in sect. \ref{S:Numr}. These calculations used the Heaviside related profiles derived in sect. \ref{S:Hp}. We note that the following characteristic timescales
are strictly valid in the framework of {\it infinitely extended} diffusive zones but are used to discuss the effect of a {\it spatially limited} diffusive zones. However, we see in sect. \ref{S:Numr} that these approximations lead to correct energy spectrum features, except at the highest energies.

\subsubsection{General statements about turbulence parameters}
\label{S:cons}

\cite{Pohletal05} discussed various possible downstream relaxation processes.  
First, the non-linear Kolmogorov damping produces a relaxation length
$\ell_{\rm d}(k) \propto k^{(\beta-3)/2}$. Each  turbulence mode $k$ being
in resonance with relativistic particle whose Larmor radius verifies $k\bar{r}_{\rm L}\geq 1$, we obtain
$\delta_{\rm d} = (3-\beta)/2 \ge 0$ (between 1 and 1/2 for $1\leq\beta\leq 2$). 
The two other processes considered by \cite{Pohletal05} scale as $k^{-1/2}$, namely $\delta_{\rm d} =
1/2$.  A variation range of $\delta_{\rm d}$ between 1/2 and 1 is then clearly identified. We extend it to encompass the regime $\delta_{\rm d} = 0$, a 
limiting case where relaxation lengths are spatially independent. 

What if $\delta_{\rm d}$ were negative ?  A strict lower limit to $\delta_{\rm d}$ is
given by the condition $\ell_{\rm d}(E_{\rm CR-min}) \le R_{\rm sh}$. A non-relativistic minimum resonant
energy $E_{\rm CR min} \simeq 0.1\times (\sqrt{2}-1) m_{\rm p}c^2$ seems acceptable so that
 $\delta_{\rm d}\ge \delta_{\rm d,lim}  = \ln(R_{\rm sh}/\ell_{\rm d,M}) /\ln(E_{\rm CR-min}/E_{\rm CR-max})$. The lower limit $\delta_{\rm d,lim}$ has
typical values of between -0.3 and -0.2 when identifying $\ell_{\rm d,M}$ with the size of the X-ray filament. Relaxation regimes with $\delta_{\rm d} < 0 $ do not necessary 
correspond to any known damping process but have some interesting properties, in particular concerning the radio filaments. 

\subsubsection{The dominant loss mechanism}
Comparing typical energy loss timescales is a useful tool for determining whether or not diffusive particle losses can affect the energy  spectrum of relativistic particles. Assuming that turbulence relaxation follows a Heaviside prescription, we can express these timescales by assuming a constant downstream magnetic field on the relaxation length $\ell_{\rm d}$ relative to a particle of energy $E$. 

Four timescales are relevant to set the maximum particle energy in a relaxed and compressed turbulence: 
\begin{enumerate}
\item The acceleration timescale is given by
\beq
\label{Eq:taucc_num}
t_{\rm acc}(E) \simeq [7\ \rm{yrs}] \times \bar{\rho}_{\rm M} \times \bar{y}(r)  {\rm K}(\beta,r) 
\times {\lambda_{\rm max,d-2} \over V_{\rm sh,4}^2} \times \left({\rho_{\rm d}(E) \over \rho_{\rm M}}\right)^{2-\beta} \ ,
\eeq 
where ${\rm K}(\beta,r)= q(\beta) \times (H(\beta,r)/r+1)$ and the maximum wavelength of the downstream turbulence is expressed in units of $10^{-2} \ \rm{pc}$.

\item The advection timescale, i.e., the time required for a particle to travel over a
  distance $\ell_{\rm d}$ while being advected with the  downstream flow, is given by
  \beq
\label{Eq:tadv}
t_{\rm adv}(E) = {\ell_{\rm d}(E) \over V_{\rm d}} \simeq  [4
  \ \rm{yrs}]\times \bar{r} \ V_{\rm sh,4}^{-1} \ell_{\rm d-2}(E) 
\ .  \eeq
\item The diffusive timescale, i.e., the time required for a particle to travel 
  over a distance $\ell_{\rm d}$ in a diffusive motion. \footnote{The factor 6 in the denominator of
Eq.(\ref{Eq:tdiff}), which appears to be the random walk along the radius of a sphere
is composed  of 3 independent random walks along each cartesian
coordinates}
\beq
\label{Eq:tdiff}
t_{\rm diff}(E) = {\ell_{\rm d}(E)^2 \over 6 D_{\rm d}(E)} \simeq [0.3\, \rm{yrs}] \times {\ell_{\rm d-2}(E)^2 
\over q(\beta)  \bar{\rho}_{\rm M} \lambda_{\rm max,d-2}} \times \left({\rho_{\rm d}(E) \over \rho_{\rm M}}\right)^{\beta-2} \ . 
\eeq 

\item{The synchrotron loss timescale 
\beq
\label{Eq:tsyn}
t_{\rm syn}(E) \simeq [1.25 \times 10^3 \ \rm{yrs}] \times E_{\rm TeV}^{-1} \times B_{\rm
  d-4}^{-2} \times f_{\rm sync} \ ,
\eeq 
where the parameter ${\rm f}_{\rm sync}$ is represented by $(H(\beta,r) + r) / (H(\beta,r)/r_{\rm B}^2+r) $. This expression takes into account the mean residence time both in the upstream and downstream medium.}
\end{enumerate}

The maximum electron energy is given by the equality $t_{\rm acc}(E_{\rm e-max})=t_{\rm loss}(E_{\rm e-max})$, where $t_{\rm loss}$
is the shortest of the synchrotron, advective, and diffusive timescales.
When X-ray filaments are controlled by the radiative losses, we have $t_{\rm loss}=t_{\rm syn}$.
In the case of escape losses being the most significant losses for filaments, we then have $t_{\rm loss} = {\rm min}(t_{\rm diff}, t_{\rm adv})$.
It can be seen that particles of energy close to $E_{\rm e-max}$, diffusive losses are always dominant compared to the advection losses, hence
$t_{\rm loss}(E_{\rm e-max})=t_{\rm diff}(E_{\rm e-max})$. We note that the downstream residence time $t_{\rm res,d}\simeq (V_{\rm d}/c) t_{\rm acc}$ (during one Fermi cycle) should not be compared with diffusive or advective timescales because only particles returning to the shock experience a full Fermi cycle. Performing this comparison would lead to a maximum particle energy $E_{\rm e-max}$ much higher than values obtained in the context of our numerical simulations.
\subsubsection{Conditions for an efficient particle acceleration}
\label{S:Cpa}
For relaxation-dominated  filaments, the ratios 
of the acceleration timescale (Eq. \ref{Eq:taucc_num}) to either the diffusive (Eq. \ref{Eq:tdiff}) 
and to the advective (Eq. \ref{Eq:tadv}) timescales vary as $E^{2(2-\beta-\delta_{\rm d})}$ 
and $E^{(2-\beta-\delta_{\rm d})}$, respectively. Two different regimes are now discussed.

\begin{table*}[ht]
\begin{center}
\begin{tabular}{c}
\begin{tabular}{|c|||c|c|c|c|}
 \hline  SNR&
\begin{tabular}{c}  $B_{d-diff}$ \\
\hline       \begin{tabular}{c} $q(1)=1$ 
       \end{tabular}
\end{tabular}  &
\begin{tabular}{c}
       $B_{d-diff}$ \\
\hline       \begin{tabular}{c} $q(1)=1$ 
      \end{tabular}   
\end{tabular} &
\begin{tabular}{c}  $B_{d-lim}/B_{d-diff}$ \\
\hline       \begin{tabular}{c} $q(1)=1$ 
       \end{tabular}
\end{tabular}  &
\begin{tabular}{c}
       $B_{d-lim}/B_{d-diff}$ \\
\hline       \begin{tabular}{c} $q(1)=16$ 
      \end{tabular}   
\end{tabular} \\
\hline Cas A & \begin{tabular}{c}
           $311$ 
          \end{tabular} &
          \begin{tabular}{c}
          $ 394 $ 
	  \end{tabular} &
\begin{tabular}{c}
           $ 2.3 $ 
          \end{tabular} &	  
\begin{tabular}{c}
           $ 0.4 $ 
          \end{tabular}  \\
\hline Kepler & \begin{tabular}{c}
           $ 220 $ 
          \end{tabular} &
          \begin{tabular}{c}
          $ 293 $ 
	  \end{tabular} &
\begin{tabular}{c}
           $ 3 $ 
          \end{tabular} &	  
\begin{tabular}{c}
           $ 0.5 $ 
          \end{tabular}  	\\
\hline Tycho & \begin{tabular}{c}
           $ 210 $ 
          \end{tabular} &
          \begin{tabular}{c}
          $ 333 $ 
	  \end{tabular} &
\begin{tabular}{c}
           $ 5.1 $ 
          \end{tabular} &	  
\begin{tabular}{c}
           $ 0.8 $ 
          \end{tabular} 	\\
\hline SN 1006 & \begin{tabular}{c}
           $ 174$ 
          \end{tabular} &
          \begin{tabular}{c}
          $ 189 $ 
	  \end{tabular} &
\begin{tabular}{c}
           $ 1.2 $ 
          \end{tabular} &	  
\begin{tabular}{c}
           $ 0.2 $ 
          \end{tabular}  	\\
\hline G347.3-0.5 & \begin{tabular}{c}
           $ 164$ 
          \end{tabular} &
          \begin{tabular}{c}
          $ 183 $ 
	  \end{tabular} &
\begin{tabular}{c}
           $ 0.95 $ 
          \end{tabular} &	  
\begin{tabular}{c}
           $ 0.15 $ 
          \end{tabular} \\
 \hline  & $\delta_{\rm d}=0$ & $\delta_{\rm d}=1/2$ &  $\forall \delta_{\rm d}$ & $\forall \delta_{\rm d} $\\ \hline
\end{tabular} 
\end{tabular}
\end{center}
\caption{Table presenting analytical estimates of the downstream magnetic
  field value in the context of {\it diffusive-loss-dominated} SNRs rims.
  The SNR rim observed parameters are the same as in  \cite{Parizotetal06}
  and the shock compression ratios are $r_{\rm tot}=r_{\rm sub}=4$.} 
\label{T:tab2}
\end{table*}
\vspace{0.2cm}

\noindent \underline{$2-\delta_{\rm d}-\beta > 0$}: Once $E \le E_{\rm e-max}$, 
the various timescales order as $t_{\rm acc} \le t_{\rm diff}$  and
$t_{\rm acc} \le t_{\rm adv}$: the acceleration process can occur
without noticeable losses and thus a particle energy spectrum behaves as a power law. 
We note that for energy lower than  $E_{\rm adv}$, advection
losses become dominant compared to the diffusive losses. Formally, we derive this energy limit by setting
 $t_{\rm acc}(E_{\rm e-max})=t_{\rm diff}(E_{\rm e-max})$, which leads to 
\beq 
\label{Eq:Eadv}
E_{\rm adv} = E_{\rm e-max} \times \left({g(r) \over 6}\right)^{1/2(2-\delta_{\rm d}-\beta)} \ ,
\eeq 
where $g(r)=3/(r-1) \times (H(\beta,r)/r+1)$. We hereafter note that $\bar{g}(r)=g(r)/g(4)$.
\vspace{0.2cm}

\noindent \underline{$2-\delta_{\rm d}-\beta < 0$}: In this case, the ratio
of the diffusive to advective timescales is always lower than unity, i.e., 
diffusive losses dominate at all energies. Once $E \le E_{\rm
e-max}$, downstream escapes limit the shock acceleration process
considerably as the acceleration time becomes longer than $t_{\rm diff}$ as energy decreases. 
The same conclusion can be reached from a close examination of the particle distribution given in Eq.(\ref{Eq:OS}). 
The term $z_{\rm d}= u_{\rm d} \ell_{\rm d}/D_{\rm d}$ is proportional to $(t_{\rm diff}/t_{\rm acc})^{1/2}$, and $2-\beta-\delta_{\rm d} < 0$
leads to $z_{\rm d}$ tending toward zero. The particle energy spectrum then steepens at low energy, which is obviously in complete disagreement with the Fermi acceleration scenario.

Hence, efficient Fermi  acceleration is only possible if $2-\delta_{\rm d}-\beta \ge 0$. 
For instance, an energy independent relaxation
length $\delta_{\rm d} = 0$ (as well as $\delta_{\rm d} < 0$) verifies this criterion for all diffusion regimes. In the case of a Kolmogorov type
non-linear turbulence damping, the supplementary relation $\delta_{\rm d} =
(3-\beta)/2$ imposes $\beta \le 1$, which means that only the Bohm regime can fulfil the
previous condition (we see in sect. \ref{S:Numr} that particle acceleration is inefficient in that case). 
In the context of Alfv\'en and magnetosonic cascades, Kolmogorov turbulence regime ($\beta=5/3$) is the sole regime failing to verify the previous condition.

\subsubsection{Magnetic field limits in a relaxed-compressed turbulence}
\label{S:Mfrel}

In the context of X-ray filaments controlled by the downstream turbulence damping, we can link the size of the filament, noted $\Delta R_X$, to the maximal
relaxation length $\ell_{\rm d,M}$ as $\ell_{\rm d,M} = \Delta R_X(E_{\rm e-max}/E_{\rm e-obs})^{\delta_d}= C(\delta_d) \Delta R_X$.
\footnote{The dependence of $\ell_{\rm d}$ on the wavelength $\lambda$ is a priori
valid only up to $\lambda_{\rm max} \simeq R_{\rm L}(E_{\rm CR-max})$ and, we should strictly not expect  the scaling of $\ell_{\rm d}$ to extend
beyond $\lambda_{\rm max}$. Above $\lambda_{\rm max}$, the diffusion coefficient increases as $R_{\rm L}^2$
and particle acceleration continues to proceed beyond $E_{\rm CR-max}$ but the number of particle accelerated and 
the turbulence energy density both rapidly drop. For this reason, we consider $\delta_{\rm d}$ to be controlled by 
the kernel of the damping rate above $E_{\rm CR-max}$ as, e.g., in the case of the Kolmogorov damping $\delta_{\rm d} = 3/2$ 
in this energy regime.} 
The energy $E_{\rm e-obs}$ is the energy of particles emitting in the $4-6$ keV band and this value depends on the local value of the total magnetic field. 

A downstream magnetic field estimation $B_{\rm d, diff}$ can be obtained from the dynamics of the electrons by requiring that $t_{\rm acc}(E_{\rm e-max})=t_{\rm diff}(E_{\rm e-max})$ using the previous relation between $\ell_{\rm d}$ and $\Delta R_X$. In the context of the Bohm diffusion, one obtains  
\beq
\label{Eq:Btacc}
B_{\rm d,-4, diff} \simeq 3.7 \times q(\beta=1)^{2/3} \times \left({E_{\rm \gamma-cut,keV}^{1/2} \bar{y}(r)^{1/2} \over \Delta R_{\rm X,-2}
C(\delta_{\rm d}) V_{\rm d,3}}\right)^{2/3} \ ,
\eeq
where $V_{\rm d,3}=V_{\rm d}/10^3 \ \rm{km/s}$ and again $\bar{g}(r)=g(r)/g(r=4)$.  If $\beta > 1$, the derivation of the magnetic field amplitude is more cumbersome. 

The determination of the SNR X-ray filaments are dominated wether either by the relaxation of the downstream magnetic turbulence or synchrotron losses  is provided by the condition $t_{\rm diff}(E_{\rm
  max}) = t_{\rm syn}(E_{\rm max}) = t_{\rm acc}(E_{\rm max})$. The corresponding limit value of the magnetic field is (again in case of Bohm diffusion)
 \beq
\label{Eq:BUL}
B_{\rm d,-4,lim} \simeq 8.9 \times \left({V_{\rm sh,4} \over \bar{r} \bar{g}(r)^{1/2}} \times 
{E_{\rm \gamma-cut,keV}^{-1/2} \over  C(\delta_{\rm d}) \Delta R_{\rm x,-2}} \times {\rm f}_{\rm sync} 
\right)^{2/3} \ .
\eeq 
To ensure that SNR X-ray filaments dominated by the relaxation of the magnetic field, it is compulsory to have $B_{\rm d,diff} < B_{\rm d,lim}$.    
 The factor $q(\beta)$  was isolated in Eq. (\ref{Eq:Btacc}) to show that no solution is then possible if $q(\beta=1) \gg 1$. In other words, a diffusion coefficient close to the Bohm value is required to allow the relaxation of the turbulence to control the size of the filaments. We have also to keep in mind that the downstream magnetic field amplitude has to be consistent with the aforementioned assumption that an amplification upstream has occurred, namely  $B_{\rm d} \gg B_{\rm ISM}$. 

Concerning cosmic rays of energy $E \sim E_{\rm CR-max}$, the downstream diffusive losses will dominate if particles cannot escape from
the upstream region into the ISM. This imposes a constraint on the magnetic field amplitude at the shock obtained from Eq.(\ref{Eq:DRV}). Indeed, upstream escape losses are dominant  if $t_{\rm acc}(E_{\rm CR-max}) < t_{\rm diff}(E_{\rm CR max})$, when we assume that $\ell_{\rm d}(E_{\rm CR-max})= \Delta R_{\rm X} (E_{\rm CR-max}/E_{\rm e-obs})^{\delta_{\rm d}}$.
\begin{itemize}
\item{In  the case $\delta_{\rm d} = 0$, downstream diffusive spatial escape downstream always controls the maximum CR energy. }
\item{For $\delta_{\rm d} \neq 0$, the previous condition leads to an upper limit to the downstream magnetic field, noted $B_{\rm d, esc}$.
Hence if $B_{\rm d} \ge B_{\rm d,esc}$ the CR maximum energy will be fixed by the upstream escape losses and conversely if $B_{\rm d} \le B_{\rm d,esc}$ the 
CR maximum energy will be set by the downstream escape losses.}
\end{itemize}
  The downstream magnetic field  then has to 
fulfil $B_{\rm d,diff} \le {\rm min}(B_{\rm d,esc}, B_{\rm d, lim})$ to ensure that the downstream turbulence relaxation be the controlling process of the energy cut-off of relativistic particles. Applying the previous conditions to our SNR sample, we always find that $B_{\rm d,lim} < B_{\rm d, esc}$. This means that an intermediary regime may exist where electrons lose their energy through radiative losses while cosmic rays cut-off is set by 
downstream diffusive losses. Of course, if the magnetic amplification process is efficient enough to generate higher turbulent magnetic field amplitude then upstream losses will take over. 

Table (\ref{T:tab2}) displays the values of $B_{\rm d, lim}$ and $B_{\rm d,diff}$ related to our SNR sample. The Kolmogorov regime was discarded as it does not produce any efficient acceleration as we see in sect \ref{S:Numr}. 
We show that for SN 1006 and G347.3-0.5, not much room is effectively left for the case of magnetic relaxation-controlled filaments. This result seems 
rather robust as a variation of the shock velocity by a factor of $40\%$, or a variation in the synchrotron cut-off by a factor of $2$ does 
not lead to any variation in the magnetic field greater than $25\%$. However, a variation in the filament width by a factor of $2$ would imply a variation 
in the magnetic field by a factor of $60\%$, which may slightly modify the previous conclusion.  Quite generally, the maximum magnetic field amplitude is found to be in the range $\sim 200-300 \ {\rm \mu Gauss}$. \\
To summarise we find that if downstream magnetic relaxation controls the features of the SNRs X-ray filaments, a Bohm-like diffusion regime is likely to occur and the particle diffusion coefficient normalization factor  $q(\beta =1)$ has to be quite close to unity, i.e., the diffusion regime has to be close to a genuine Bohm diffusion regime. In this context, we show that only a fraction of our SNR sample to meet these conditions, namely the young ones. Using the various observational constraints related to the older SNRs (SN1006 and G347.3-0.5), we have shown that the X-ray filaments existing in these objects are likely to be ruled by radiative losses associated with synchrotron emission.  

\subsubsection{Radio filaments}
The energy of the radio electrons is typically four order of magnitude
below that of the X-ray emitting electrons, i.e.,
\[
E_{\rm eobs,R} \simeq [1.5 \rm{GeV}] \ B_{\rm d,-4}^{-1/2} E_{\rm \gamma-obs-GHz}^{1/2} \ , 
\] 
where $E_{\rm  \gamma-obs-GHz}$ is the energy of the radio electrons emitting in the
GHz band. Using both Eqs. (\ref{Eq:tsyn}) and (\ref{Eq:tadv}),  one
can easily check that the synchrotron loss timescale at $E_{\rm eobs,R}$ is
always longer than the advective loss timescale, unless $\delta_{\rm d}$
is lower than typical values of the order of $-0.5$, a value always lower
than $\delta_{\rm d,lim}$. If $ \delta_{\rm
  d-lim} \le \delta_{\rm d} \le 0$, the small turbulence scales relax on distances longer than
$\Delta R_{\rm X}$. This very particular case would produce radio filaments
larger than the size of X-ray filaments inferred from the Chandra
observations. In contrast, the regime $\delta_{\rm d} \ge 0$ would allow the largest fluctuating
scales controlling the size of the  radio filaments. In this case, the radio
filaments are expected to be of the order of $\Delta R_{\rm X}$ (see
\cite{Cassamchenaietal07}).

\begin{figure}
  \resizebox{\hsize}{!}{\includegraphics{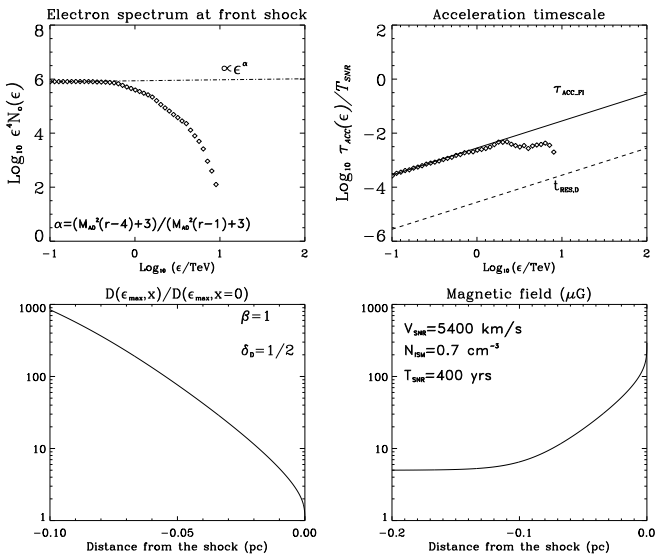}}
   \caption{Energy spectrum of relativistic electrons at the shock front given by MHD-SDE
     simulations in the conditions of the Kepler SNR (see Fig.\ref{F:F31} for details).}
   \label{F:Fig41}
\end{figure}

\subsection{Numerical simulations}
\label{S:Numr}
We performed MHD-SDE simulations by taking into account all previous
settings, namely the downstream magnetic field relaxation, the stochastic reacceleration,
and the radiative losses for the electrons.  In the following paragraphs, we discuss the physical agreement between
assuming magnetic field relaxation to control the X-ray filaments  and the actual results coming from 
the computation of relativistic electrons acceleration. 

\subsubsection{Downstream magnetic Kolmogorov damping}

When non-linear Kolmogorov damping occurs in the downstream medium of the shock, we have seen in the previous 
sections that two conditions have to be fulfilled to reproduce both the appropriate energy cut-off and the correct size of the observed
X-ray filament. These two conditions can be expressed as: having the correct downstream magnetic field given by Eq.(\ref{Eq:Btacc})
(to ensure that the electron energy cut-off is consistent with the observations) and having the typical magnetic relaxation length $x_{O-K}$  
(see Eq. \ref{Eq:x0K}) that is similar in the size of the X-ray filament. In the non-linear Kolmogorov regime, the only diffusion regime able to 
provide to an efficient particle acceleration is the Bohm diffusion regime, where the relaxation energy index $\delta_d=1$. Inserting, for the Kepler SNR,  this value into Eq.(\ref{Eq:Btacc}) leads to a downstream magnetic field of $B_{\rm d} \simeq 390 \mu G$ and a relaxation of $x_{\rm 0-K}\simeq 0.39$ pc. The relaxation size is clearly too large to provide an X-ray filament, whose thickness is inferred to be of the order of $10^{-2}$ pc from X-ray observations. Applying the same reasoning to the other SNRs leads to a similar conclusion: having both the appropriate electron energy cut-off and X-ray filament size is incompatible with a non-linear Kolmogorov occurring in the downstream medium of the SNR shock. The only way to overcome this conclusion would be to have the factor $\sigma=\ln(k_{\rm max}/k_{\rm min})$ to be much smaller than expected (see Eq.\ref{Eq:x0K}). Anyway, having $\sigma$ so low would mean that the range of particle energy able to resonate with turbulence mode would be so narrow that it would not be able to provide any significant acceleration. This explains why our result differs from the conclusion drawn by \cite{Pohletal05}. It seems then that it is very unlikely that non-linear Kolmogorov damping, which is a slower process than Alfv\'en/ fast magnetosonic cascade, occurs in the downstream medium of SNR shocks.

\subsubsection{Alfv\`enic-fast magnetosonic mode damping}

\begin{figure}
  \resizebox{\hsize}{!}{\includegraphics{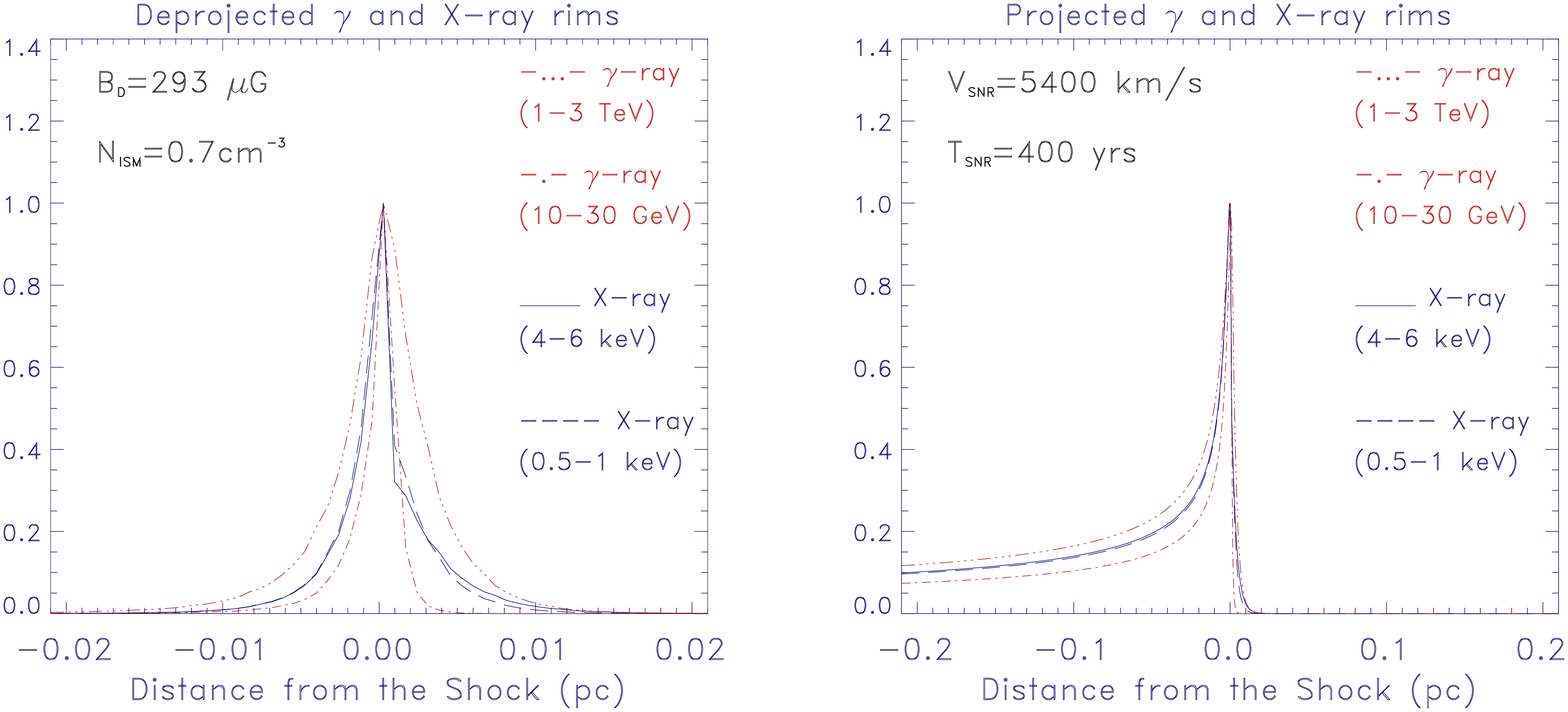}}
   \caption{ The unprojected and projected X-ray and $\gamma$-ray rims in the conditions of the Kepler SNR in the case of an exponential relaxation profile. For clarity, both X- and $\gamma$-ray rims have been normalised to one.}
   \label{F:Fig42}
\end{figure}

In the context of Alfv\'enic-fast magnetosonic turbulence relaxation, the typical relaxation length $x_{\rm 0-A}$ is shorter than $x_{\rm 0-K}$. Compiling the aforementioned necessary conditions  to reproduce accurately an X-ray filament in the SNR environment, we obtain a typical $x_{\rm 0-A}$ of the order of $10^{-2}$ pc when using magnetic field values provided by Table(\ref{T:tab1}). This means that the Alfv\'enic-fast magnetosonic modes damping is a plausible candidate to explain the presence of SNRs X-ray filaments. To sustain this conclusion, we performed, in the context of the Kepler SNR, MHD-SDE simulations designed to reproduce the dynamics of relativistic electrons and the associated X-ray and $\gamma$-ray emission maps.
In figures \ref{F:Fig41} and \ref{F:Fig42}, we display the particle distribution at the shock front and the X- and $\gamma$-ray filaments respectively. All simulations were performed in the Bohm regime. In that case, $2-\delta_{\rm d}-\beta = 1/2 > 0$. In each cases the magnetic field is damped in the downstream medium following an exponential relaxation as in Alfv\`enic-fast magnetsonic modes damping. Bohm regime in downstream region has been assumed. The dashed-line shows the stationary solution found in \cite{Marcowithetal06}, which includes particle reacceleration in the Fermi cycle. In the upper right panel the acceleration (only the regular Fermi acceleration), and the diffusive and downstream residence timescales are displayed using solid and dashed lines. Diamonds represent for our numerical calculation of the acceleration timescale, which is in agreement with the theoretical estimation.  We also display in the two lower panels the spatial dependence of the
     diffusion coefficient at the maximum electron energy (lower left) and the magnetic profile in the downstream medium at $t=400$ years (lower right).

Several obvious differences appear in both figures \ref{F:Fig41} and \ref{F:Fig42} with respect to the simple advection case presented in figures \ref{F:F31} and \ref{F:F32}. First, as stated in sect. \ref{S:Arc} the normalization of the diffusion coefficient $q(\beta)$ has to be close to one. Even in this case, the maximum particle energy is limited to values close to ten TeV (for parameters associated with the Kepler SNR).  One of the necessary conditions to fit the observed size of the X-ray rim, namely   $x_{\rm 0-A}\sim \Delta R_{\rm X}$,  
produces an increase in the diffusion coefficient by a factor of a few tens above the typical diffusion length, and consequently low maximal energies for both electrons and cosmic rays. The X- and $\gamma$-ray filaments also exhibit some different features in the case of an Alfv\`enic-like relaxed turbulence.
The low energy particles producing the synchrotron photons in the interval 0.5-1 keV and the $\gamma$-ray photons in the 10-30 GeV band, respectively, do extend to shorter distances behind the shock (electrons having energy $\sim 1$ TeV). This can be understood by the effect of the resonant component of the magnetic field $b$ in Eq.(\ref{Eq:nus}). At a given downstream location,  particles with  energies $E < E_{\rm max}$ do interact with a lower number of modes than in the advected case. This effect is caused by high wave number modes relaxing over shorter distances than lower wave number modes within the same turbulence spectrum. Compared to the advected case, more low energy particles experiencing diffusive losses are lost than at highest energies (which are also subject to diffusive losses).  Particles of energy around a few tens to hundreds of GeV are then confined to closer to the shock and do not experience strong magnetic field variation: the standard shock solution is then recovered in this domain. We verified that the shock synchrotron spectrum cuts off at an energy close to one keV. 

\begin{figure}
  \resizebox{\hsize}{!}{\includegraphics{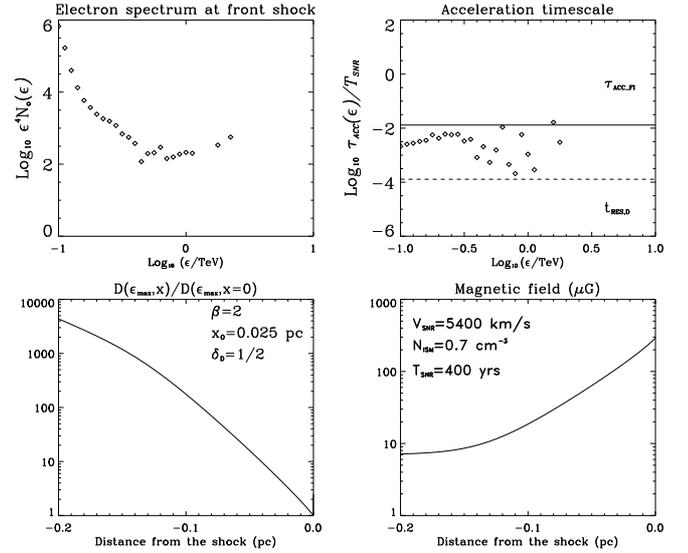}}
   \caption{Same case as treated in Fig.\ref{F:Fig41} but with $\beta = 2$. Here, massive diffusive losses are occurring since $2-\delta_d-\beta < 0$ and thus no significant acceleration is observed.}
   \label{F:Fig43}
\end{figure}

We also tested the solution in the case $\beta =2$, i.e., $2-\delta_{\rm d}-\beta = -1/2 < 0$. No significant particle acceleration has been found as diffusive losses dominate at low energy (see Fig.\ref{F:Fig43}). The numerical acceleration timescale is also found to be shorter than the theoretical estimation which is consistent with particles only returning quickly to the upstream medium after entering the downstream region, are able to avoid massive diffusive losses. These simulations confirm the conclusions drawn in sect. \ref{S:Cpa}. 

\subsubsection{Solutions for turbulent dynamo amplification}
The coherence length of the downstream turbulence entering the evaluation of $\ell_{\rm ampl}$ in sect. \ref{S:TDD} cannot be longer than the X-ray filament
width, otherwise the condition about the maximum CR diffusion coefficient upstream given by Eq.(\ref{Eq:DRV}) would not be satisfied. This means that if a magnetic dynamo
operates downstream, then the growth scale length is $< \Delta R_{X}$. The growing modes are restricted mostly to large scales, i.e., to wave numbers close to $k_{\rm min}$.
They are considered for the particles to contribute to the mean magnetic field. The rapid increase in the magnetic field downstream to values close to equipartition 
produces enhanced radiative losses and thus much thinner filaments.  We checked the effect by performing simulations in which we added a mean magnetic field downstream of values close to a few mGauss. 

\section{Discussion and summary}
Young SNRs are strong particle accelerators, as illustrated by the presence of thin X-ray filaments. In these astrophysical objects, the X-ray emission is produced by synchrotron radiation, involving particles whose maximal energy is higher than tens of TeV and magnetic field strengths behind the shock of a few hundred $\mu$Gauss \citep{Parizotetal06}. This work has extended the study undertaken by \cite{Parizotetal06} of the physical properties of both the turbulence and transport coefficients in the same sample of five young SNR.  We have  included the turbulence compression at the shock front, the possibility of particle reacceleration in the downstream region of the shock, and the relaxation of the magnetic fluctuations downstream \citep{Pohletal05}. We have also described the generation of magnetic fluctuations in the shock precursor for the two regimes of the streaming instability \citep{Pelletieretal06}. This work has been developed in the same framework as \cite{Lagage83} but adapted to the case of amplified magnetic fields around SNR, although the maximum CR energy has not been fully investigated here. We have 
developed a numerical scheme based on the coupling between the equations of magnetohydrodynamics and a kinetic scheme handling the calculation of the electron particle distribution function. The scheme involves a set of stochastic differential equations (SDE) described elsewhere \citep{Cass03,Cass05}. The SDEs have been adapted to account for the discontinuity in the diffusion coefficients properly using a skew Brownian motion (see also \citet{Zhan00}). The following conclusions can be made: 

\begin{enumerate}

\item The compression of turbulent scales at the shock front does not deeply modify the efficiency of shock acceleration. The conclusions addressed by \cite{Parizotetal06} are found to be robust in the case of a downstream, advected, magnetic field, young SNRs exhibiting X-ray filaments do accelerate particles to at most PeV energies.

\item For the various regimes of streaming instability occurring in the shock precursor, the SNRs contained in our sample are expected to generate magnetic fields up to a few hundred $\mu$Gauss. For shock velocities of a few hundred thousand km/s, the level of fluctuations tends to be shared by the non-resonant and the resonant regimes. The resonant modes may contribute to some particle reacceleration downstream. However, the amount of reacceleration cannot be too large, otherwise the shock particle spectrum would be harder and the X-ray filament width would be larger than observed. This provides an observational constraint of the number of resonant modes present downstream of the shock front. The fate of non-resonant modes generated upstream still requires consideration.

\item We have presented calculations of the projected and deprojected X- and $\gamma$-ray filaments, each one in two specific wavebands. For the separation between  the X and $\gamma$-ray peak emission is found to be far below any $\gamma$-ray mission resolution capabilities when observing young SNR, some detailed observations could be undertaken for more extended objects such as Vela Junior. 

\item For relaxed turbulence occurring in the downstream region, our conclusions are the following:
\begin{itemize}
\item When the magnetic relaxation scale varies as $\ell_{\rm d}(k) \propto k^{-\delta_{\rm d}}$,  a magnetic turbulence (whose power-law index is $\beta$) is able to provide suitable conditions giving rise to  an efficient particle acceleration if
$2 -\delta_{\rm d} -\beta > 0$. 

\item We have tested several relaxation processes obtaining various values of $\delta_{\rm d}$. When Kolmogorov damping occurs in a Bohm diffusion regime,  it appears unlikely to produce strong acceleration in the framework
of relaxation limited filaments when accounting for the complete dynamics of the turbulent spectrum. On the other hand, the Alfv\'en and fast magneto-sonic cascades provide suitable conditions giving birth to particle acceleration while being able to match all observational features of X-ray filaments. In this context, we have found that the maximum energy particle (both for electrons and cosmic rays) cannot be  much higher than a few tens of TeV.

\item The magnetic field strengths downstream of the shock cannot be much higher than $200-300\mu$ Gauss, otherwise radiative losses would control the X-ray filament width. 

\item For the supernova remnants SN1006 and RXJ 1713-3946.5, none of the various turbulence relaxation processes considered in the present paper have been able to provide efficient particle acceleration and match the corresponding observational features. It seems that only the youngest SNRs ($T_{\rm SNR}< 500$ yr) of our sample may exhibit X-ray filaments controlled by downstream turbulence relaxation. 

\item The normalization (i.e., factor $q(\beta)$) of the spatial diffusion coefficient should remain close to unity to avoid massive particle diffusive losses, leading to a drop of the Fermi acceleration efficiency. A genuine Bohm diffusion regime is then required if magnetic turbulence relaxation is to occur in the downstream region of the shock.

\end{itemize}

\end{enumerate}

\begin{acknowledgement}{}
The authors thank F.Acero, E. Parizot, G. Pelletier for their valuable comments. V. Tatischeff is thanked for his careful reading of the manuscript and many suggestions. This work work has been supported by the French National Research Agency project AccFermi.
\end{acknowledgement}{}

\appendix

\section{Magnetic field profile produced by the resonant instability}
\label{S:Appa} 
The amplification factor related to the resonant instability depends on the amplification factor produced by the non-resonant 
instability (\cite{Pelletieretal06}, Eq.34) and is given by
\beq
A^2_{\rm R}(x) = \tilde{\alpha}_{\rm res} \times A_{\rm NR}(x) \times \int_1^{k_* \ell_{\rm coh}} d\ln(\bar{k}) \left(\exp(-a(x) \bar{k}^{2-\beta})
-1/e\right) \ ,
\label{Eq:Ar}
\eeq
where $\tilde{\alpha}_{\rm res}=\pi/\phi \times M_{\rm A\infty}\xi_{\rm CR} > 1$ and $k_*$ is the maximum resonant wave length at a distance $x$,
$\bar{k} = k \ell_{\rm coh}$ varies between $k_{\rm min}(=1/r_{\rm L}(E_{\rm CR-max})) \ell_{\rm coh} \simeq 1$ and $k_*(x) \ell_{\rm coh} 
\ge 1$ \footnote{As discussed in sect. \ref{S:MFnr}, we assume the same coherence length over the whole precursor.}. 
We have:
$$
a(x)={\pi \over \beta \phi} \times (V_{\rm sh}/c)\times (x/\ell_{\rm coh}) \times \eta_{\rm tot}(x) < 1 \ , 
$$
The exact integration of Eq.(\ref{Eq:Ar}) involves a difference between two exponential integral: ${\rm Ei}(-a(x)\bar{k}_*)-{\rm Ei}(-a(x))$. 
 The second term dominates when $\bar{k}_* \ge 1$, and we obtain 
\beq
\label{Eq:Arf}
A_{\rm R}(x) \propto \left[A_{\rm NR}(x) \times (-Ei(-a(x))/(2-\beta)-\ln(k_*(x))/\exp(1)) 
\right]^{1/2} \ .
\eeq
The above equation is implicit because the total magnetic field is hidden in $k_*$ and $\eta_{\rm tot}$. \\
 At distances $x \ll \ell_{\rm diff}(E_{\rm CR-max})$ where $a(x) \ll 1$, we approximate $-Ei(-a(x)) \simeq -\ln(a(x))-{\cal C}$,
${\cal C} \simeq 0.5772$ is the Euler constant. At a first approximation, within the precursor $A_{\rm R}(x)$ scales as $A(x)_{\rm NR}^{1/2}$.

\section{Derivation of the shock particle distribution function}
\label{S:Appb}
The steady-state general 1D Fokker-Planck 
equation is given by
\beq 
u\frac{\partial f}{\partial x} = \frac{\partial}{\partial x}\left(D\frac{\partial
f}{\partial x}\right) + (u_{\rm d} - u_{\rm u})\delta (x)\frac{\partial
  f}{\partial \ln p^3} \ ,
\label{Eq:gene1}
\eeq
Where the upstream medium is defined by $-\ell_{\rm u}(p)\leq x<0$ and the downstream
medium by $0< x \leq \ell_{\rm d}(p)$. The shock front is at $x = 0$. 
In this equation, we have neglected the synchrotron/turbulence generation losses since we focus on 
the particle diffusive losses. The presence of finite
extensions in both the upstream and downstream media imposed by boundary
conditions for $f$ as $f(-\ell_{\rm u},p)=0=f(\ell_{\rm d},p)$. To
determine the spatial behaviour of the $f$ function, we integrate
Eq.(\ref{Eq:gene1}) from the left boundary to $x$ in the upstream medium 
and from $x$ to the right boundary in the downstream medium, and we obtain
\begin{eqnarray}
f_{\rm u}(x,p)&=& f_{\rm S}(p)\frac{\int_{-\ell_{\rm
      u}}^x\exp(\int_{-\ell_{\rm u}}^{x'}\theta_{\rm u}(x'',p)dx'')dx'}{\int_{-\ell_{\rm
      u}}^0\exp(\int_{-\ell_{\rm u}}^x\theta_{\rm u}(x',p)dx')dx} \ , \nonumber \\
f_{\rm d}(x,p)&=& f_{\rm S}(p)\frac{\int_{x}^{\ell_{\rm d}}\exp(-\int_{x'}^{\ell_{\rm d}}{\rm
    d}(x'',p)dx'')dx'}{\int_{0}^{\ell_{\rm d}}\exp(-\int_x^{\rm \ell_{\rm d}}\theta_{\rm
    d}(x',p)dx')dx} \ ,
\label{Eq:gene2}
\end{eqnarray}
where $f_{\rm S}$ is the distribution function evaluated at the shock front
and the functions $\theta_{u/d}$ are the inverse of the effective diffusive lengths and defined to be
\beq
\label{Eq:theta}
\theta_{\rm u/d}(x,p) = \frac{u_{\rm u/d} - \frac{\partial D_{\rm u/d}}{\partial x}}{D_{\rm u/d}} \ .
\eeq
The energy flux carried by the relativistic particle has to be conserved
throughout the shock front, namely for $\upsilon \rightarrow 0$ 
\beq
\left [D\frac{\partial f}{\partial x} + u\frac{\partial f}{\partial \ln{p^3}}
\right ]_{-\upsilon}^{\upsilon} = 0 \ .
\eeq
The spatial derivatives of $f$ are evaluated using Eq.(\ref{Eq:gene2}), which produces a differential equation for $f_{\rm S}$:
\bea
\label{Eq:gene3}
\frac{d \ln f_{S}(p)}{d \ln p} & = &-\frac{3}{(u_{\rm u}-u_{\rm d})} \times \left\{\frac{D_{\rm u}(0,p) 
\exp(\int_{\rm -\ell_{\rm u}}^0 \theta_{\rm u}(x',p)dx')}{\int_{-\ell_{\rm u}}^0\exp(\int_{\rm -\ell_{\rm u}}^x
\theta_{\rm u}(x',p)dx')dx}\right. \nonumber \\
& & \left. + \frac{D_{\rm d}(0,p) \exp(-\int_0^{\rm \ell_{\rm d}} \theta_{\rm d}(x',p)dx')}{\int_0^{\ell_{\rm d}}
\exp(-\int_x^{\rm \ell_{\rm d}}\theta_{\rm d}(x',p)dx')dx}\right\}.
\eea

\section{Particle acceleration and multi-scale simulations}
\label{S:Appc}
This section presents the numerical framework used to simulate both the
supernova thermal plasma evolution and the relativistic charged particles
transport. As detailed in \cite{Cass03} and \cite{Cass05}, the background
fluid and large-scale magnetic field are calculated using the
magnetohydrodynamics code VAC for {\it Versatile Advection Code}
(\cite{Toth96a}). The simulations are performed using a 1D spherical
symmetry, where the evolution of the supra-thermal electrons and nuclei are
calculated using the stochastic differential equations (SDE) formalism
\citep{Krulls94}. The numerical description of supra-thermal particle
transport is crucially dependent on the ability of the MHD code VAC to
capture the shock structure. To obtain the sharpest shock front
possible, we used the TVD-MUSCL scheme coupled with a Roe-type approximate
Riemann solver (\cite{Toth96b}).\\ 

Section \ref{S:SNR} briefly reports on the MHD-SDE schemes used to model a 1D spherical SN remnant expansion.
In particular, sects \ref{S:Scheme} and \ref{S:SDE2D} discuss at length the stochastic differential
Euler schemes with spatially dependent diffusion coefficients and their application to the diffusive shock acceleration
problem. Section \ref{S:KinMHD} describes the shock capturing procedure that efficiently couple the MHD and SDE
schemes.

\subsection{Supernova remnant modelling}
\label{S:SNR}
The time evolution of the thermal magnetised plasma is fully controlled by
the MHD equations providing mass, momentum, and  energy conservation as well
as electromagnetic field induction, namely
\begin{eqnarray}
\frac{\partial \rho}{\partial t}&+&\nabla \cdot (\rho \UU)=0
\ ,\nonumber\\ \frac{\partial (\rho \UU)}{\partial t}&+& \nabla \cdot [
  \rho \UU \UU + p_{tot} I- \BB \BB/\mu_o]=0\ ,\\ \frac{\partial e}{\partial
  t}&+& \nabla \cdot \left( e\UU + p_{tot} \UU - \UU \cdot \frac{\BB
  \BB}{\mu_o} \right)= 0 \ ,\nonumber \\  e &=& \frac{\rho \UU^2}{2} +
\frac{\BB^2}{2\mu_o}+\frac{P}{\gamma -1}\nonumber \\    \frac{\partial
  \BB}{\partial t}&+& \nabla \cdot (\UU \BB-\BB \UU)= 0\nonumber
\label{MHD1}
\end{eqnarray}
The density $\rho$, velocity $\UU$, total energy $e$ and magnetic field
$\BB$ are set by the initial conditions as a 1D spherically symetric SNR
blast-wave described by \cite{True99}. We assumed a uniform SNR and added a small 
contribution of the magnetic field. The resulting SNR
MHD simulation starts for ($V_{\theta},V_{\phi}=0$) with the parameters
\[ \rho = \left\{ \begin{array}{cc}
       3M_{\rm SNR}/\rho_{\infty}4\pi V_{\rm SNR}^3T_{\rm SNR}^3 & \ ,
       R<V_{\rm SNR}T_{\rm SNR}\\ 1 & \ , R>V_{\rm SNR}T_{\rm SNR} \end{array}\right. \]
\[ V_{\rm R} = \left\{ \begin{array}{cc}
       R/V_{\rm SNR}T_{\rm SNR} & \  , R < V_{\rm SNR}T_{\rm SNR}\\ 0 & \ ,
       R>V_{\rm SNR}T_{\rm SNR} \end{array}\right. \]  
For each run, the physical quantities entering the problem are normalised by the known mass ejected
$M_{\rm SNR}$, the age of the  SNR $T_{\rm SNR}$, the mechanical energy of the 
explosion $E_{\rm inj}$, and the velocity of the blast wave
$V_{\rm NR}$. We set the thermal pressure to a small value compared to the
kinetic energy of the SNR (typically $10^{-3}$ times), since its role in
the wave propagation is minimal. The magnetic field advected along the
flow is also believed to be very ineffective in the wave propagation but
its role in the supra-thermal particles transport process is
important. The magnetic field is thus prescribed with an amplitude similar to
its warm interstellar medium value, e.g., $B_{\theta}\simeq 5\mu G$.\\ 
To test the ability of our simulation to model the propagation of 
SNR shock, we simulated the long-term evolution of a SNR blast wave corresponding 
to the previous initial set-up where we defined the SNR parameter 
to $M_{\rm SNR}=6M_{\odot}$, $T_{\rm SNR}=200\rm{yr}$, $E_{\rm inj}=10^{51} \rm{ergs}$ 
and $V_{\rm SNR}=5000 \rm{km/s}$. The results were found to reproduce the corresponding analytical
solution in \cite{True99} quite accurately. In particular, both the free expansion and Sedov self-similar
regimes were obtained, the transition regime occurring at the expected Sedov time for this simulation of $T_{\rm SEDOV}=1.1\rm{kyr}$. 

\subsection{Kinetic approach}
\label{S:Scheme}
The transport of relativistic particles (with velocities much larger than
the fluid speed) near the shock front is governed by a Fokker-Planck
equation when these particles resonate with the turbulence and
enter a diffusion regime. The related kinetic equation is  
\begin{eqnarray}
 \frac{\p F}{\p t}= &-&\frac{\p}{\p R}\left(F\left\{V_R+\frac{\p D_{R}}{\p
   R} +\frac{2D_{R}}{R}\right\}\right)\nonumber\\ &-&\frac{\p}{\p
   p}\left(F\left\{-\frac{p}{3}\nabla\cdot\UU+\frac{1}{p^2}\frac{\p p^2
   D_{pp}}{\p p}-a_{loss}p^2\right\}\right) \nonumber\\ &+&\frac{\p^2}{\p
   R^2}(FD_{R})+\frac{\p^2}{\p p^2}(FD_{pp}) \ ,
\label{Kine1}
\end{eqnarray}
\noindent where $F=R^2p^2f$ is related to the distribution function $f$ in terms of the
spherical radius $R$ and particle momentum  $pc=\gamma m_ec^2$. The
particle spatial diffusion regime is characterised by a diffusion
coefficient $D_R$ that depends on the turbulence spectrum. The factor $a_{\rm loss}$
stands for particle losses. \\
For electrons, the losses are produced by synchrotron
cooling. The cooling timescale $t_{syn}$ is
\begin{equation}
a_{\rm syn}= \frac{1}{t_{\rm syn}p}=\frac{6\pi m_e^2c^2}{\sigma_{\rm T} c B^2} \ .
\label{Kine2}  
\end{equation}
For protons (or ions), the losses are produced by the generation of magnetic fluctuations
and are a priori limited to the upstream medium (in the downstream flow the particle distribution
is isotropic). The cooling timescale is adapted from \cite{Marcowithetal06} their Eq.13
\begin{equation}
a_{\rm turb}= {P(p) \over p^2} \ , 
\label{Kine3}  
\end{equation}
where $P(p)$ is the rate of energy radiated by a relativistic particle
\beq
P(p) \simeq {1 \over 3} V_{\rm sc} {\partial \log(f(x)) \over \partial x} p \ .
\eeq
The scattering centre velocity is close to the local Alfv\'en velocity, i.e.,
$V_{\rm sc} \simeq V_{\rm Au}$.\\
Stochastic particle acceleration is represented by the energy diffusion
coefficient  $D_{\rm pp}=V_{\rm A}^2p^2/9D_{\rm R}$, which is related to spatial diffusion (where $V_{\rm A}$ is
the local Alfv\`en velocity).

\subsubsection{Stochastic differential equations}  
\label{S:SDE2D}
As shown by \cite{Krulls94}, this Fokker-Planck equation is equivalent to
a set of two SDEs that can be written as
\begin{eqnarray}
\frac{dR}{dt}&=&V_{\rm R}+\frac{\p D_{\rm R}}{\p
  R}+\frac{2D_{\rm R}}{R}+\frac{dW_{\rm R}}{dt}\sqrt{2D_{\rm R}}\nonumber\\
\frac{dp}{dt}&=&-\frac{p}{3}(\nabla\cdot\UU)+\frac{1}{p^2}\frac{\p
  p^2D_{\rm pp}}{\p p}-a_{\rm loss}p^2+\frac{dW_{\rm P}}{dt}\sqrt{2D_{\rm pp}} \ ,\nonumber 
\label{SDE4}
\end{eqnarray}
where $W_i$ are Wiener processes for which 
$dW_i\propto\sqrt{dt}$. Using Monte Carlo methods, it is then possible to
time-integrate the trajectories of a sample of test particles in phase
space and to reconstruct this distribution function, provided that the
number of test particles is sufficiently high.\\ A shock
discontinuity may lead, according to the MHD Rankine-Hugoniot conservation
laws, to a discontinuous magnetic field at the shock front. Depending on
the diffusion regime affecting relativistic particles, this may lead to
discontinuous diffusion coefficients that can be written $D_{\rm R}=D_{\rm R,C} +
\Delta D_{\rm R}\text{sign}(R-R_{\rm sh})$, where the first term is a continuous
function. In this case, the usual Euler schemes are no longer valid, in contrast to the studies of 
\cite{Krulls94}, \cite{Cass03}, \cite{vds04}, and \cite{Cass05}. As shown by  \cite{Zhan00}, it is
possible to overcome this problem by employing a skew Brownian motion where
an asymmetric shock crossing probability is considered. In this framework,
the spatial stochastic equation becomes
\begin{equation}
d\tilde{R} = \xi(\tilde{R})\left\{\left(V_{\rm R} + \frac{\partial
  D_{\rm R,C}}{\partial R}\right)dt + \sqrt{2D_{\rm R}}dW_{\rm R}\right\} \ ,
\label{Zh1}
\end{equation}   
where $\tilde{R}$ is related to $R$ by 
\[ \tilde{R} = \xi(R)R\ \  \text{with}\ \ \xi(R) = \left\{
\begin{array}{lr}
\varepsilon\ , & R < R_{\rm sh}\\ \frac{1}{2}\ , & R =
R_{\rm sh}\\ (1-\varepsilon)\ , & R > R_{\rm sh}\\
\end{array}\right. \ , \]
and where $\varepsilon$ is the ratio of diffusion coefficients at the
shock front, namely 
\begin{equation}
\varepsilon = \frac{D_{\rm u}(R_{\rm sh})}{D_{\rm u}(R_{\rm sh})+D_{\rm d}(R_{\rm sh})}
\ .
\end{equation}
Equation~(\ref{Zh1}) can be solved using an Euler scheme where the stochastic
variable $W_R$ is computed with Monte Carlo methods.  In contrast to the
study of \cite{Zhan00}, realistic diffusion coefficients are likely to
depend on particle energy. In this case, we have to consider the amount of
energy $\Delta\epsilon$ gained by particles during the shock crossing. The
transition probability $\varepsilon$ is then calculated depending on the
way that the shock is crossed, namely
\begin{eqnarray}
\varepsilon_{up\rightarrow down} &=& \frac{D_{\rm
    u}(R_{\rm sh},\epsilon)}{D_{\rm u}(R_{\rm sh},\epsilon)+D_{\rm
    d}(R_{\rm sh},\epsilon +\Delta\epsilon)}
\nonumber\\  \varepsilon_{down\rightarrow up} &=& \frac{D_{\rm
    u}(R_{\rm sh},\epsilon+\Delta\epsilon)}{D_{\rm u}(R_{\rm sh},\epsilon +
  \Delta\epsilon)+D_{\rm d}(R_{\rm sh},\epsilon)} \ .   
\end{eqnarray}  
We note that this skew Brownian motion approach is valid only if
shock curvature terms are negligible, i.e., $2D_{\rm R}/R \ll |V_{\rm R}+\partial
D_{\rm R}/\partial R|$. In the energy stochastic equation, the velocity
discontinuity can be numerically treated using an implicit Ricatti scheme
\citep{Marcowith99}. Basically, once the stochastic displacement $\Delta R$
is calculated, we can calculate the energy gained $\Delta\epsilon$ by a
particle originally of energy $\epsilon=pc$ during time step
$\Delta t$ following 
\begin{equation}
\frac{\epsilon+\Delta\epsilon}{\epsilon} =
\frac{\text{exp}\left(-\frac{\Delta t}{3\Delta R}\int_{R}^{R+\Delta
    R}\nabla\cdot{\bf V}dR\right)}{1 +\epsilon\ \text{exp}\left(-\frac{\Delta
    t}{3\Delta R}\int_{R}^{R+\Delta R}\nabla\cdot{\bf V}dR\right)
  \frac{\Delta t}{\Delta R}\int_{R}^{R+\Delta R} a_{\rm loss} dR} \ .
\label{Sol1}
\end{equation}       
The previous implicit calculation is valid for any diffusion regime
provided that second-order Fermi acceleration is negligible. In the
opposite case, we then have to step back into an explicit scheme to take
into account the skew Brownian motion. Following \citet{Zhan00}, the energy
gained by a particle is
\begin{equation}
\label{Eq:DE}
  \Delta\epsilon = \sqrt{2D_{\rm pp}}dW_{\rm p}-\frac{\Delta V}{3\Delta
    D_{\rm R}}\epsilon\{\Delta {\rm R}-\Delta\tilde{R}/\xi(\tilde{R})\} +
  \left(\frac{\partial  D_{\rm pp}}{\partial p}-a_{\rm loss}\right)\Delta t \ ,
\end{equation} 
where $\Delta V=V_{\rm up}(R_{\rm sh})-V_{\rm down}(R_{\rm sh})$ and $\Delta D_R=D_{\rm
  u}(R_{\rm sh})-D_{\rm d}(R_{\rm sh})$. During the time integration of MHD
equations, the SNR shock front propagates in such a way that its surface
increases with time. To take into account the increase in the
particle flux at the shock front, we continuously inject new particles
of energy $\epsilon_{\rm inj}$, so that the number of new particles is
$N_{\rm part}(t+\Delta t)-N_{\rm part}(t)\propto R_{\rm sh}^2(t)\Delta R_{\rm sh}$, where
$\Delta R_{\rm sh}$ is the shock front displacement occurring during $\Delta
t$.

\subsubsection{Kinetic description of MHD shock waves}
\label{S:KinMHD}
The SDE formalism is useful for modelling the transport of relativistic test
particles a non-relativistic  background fluid since it
provides both the spatial and energetic distribution of particles. Nevertheless
one drawback of this  method does exist: the shock thickness. The
SDE algorithm is based on the use of fluid velocity divergence to mimic
particle acceleration. The MHD code provides the velocity field at
discrete locations on the grid so that $\nabla\cdot\UU$  may be obtained
by means of linear interpolation. The most efficient MHD code cannot reproduce
shocks as sharp discontinuities but rather displays velocity and density
variations over two or three cells. This is very important for kinetic
computations since  particles with diffusion coefficients for which the
diffusive step is small compared to the MHD shock thickness will see the shock as
an adiabatic compression, leading to softer energy spectrum.\\ In previous
work (see e.g., \cite{Krulls94,Cass03}), it was shown that the SDE
formalism was able to describe accurately  the transport of particles with
diffusion coefficients greater than $\Delta X_{\rm sh}V/2$, where $\Delta
X_{\rm sh}$ is typically  the cell size in the MHD code. This constraint
greatly reduced the range of applications of this method. To
overcome this problem, we designed a SDE algorithm in which the
$\nabla\cdot\UU$ is no longer calculated locally but we instead integrate
the term $\nabla\cdot\UU dR$ in Eq.(\ref{Sol1}), where the velocity is given
as $V_{\rm u}$ or $V_{\rm d}$  depending on the shock position. In our new
approach, the MHD code now provides the shock position and the
compression ratio $r$ so that we deduce the value of the fluid by
considering the shock as infinitely thin.



\bibliographystyle{aa}

\end{document}